\documentclass[twocolumn,pre,superscriptaddress,showpacs,preprintnumbers,amsmath,amssymb]{revtex4}

\usepackage{graphicx}% Include figure files
\usepackage{dcolumn}% Align table columns on decimal point
\usepackage{bm}% bold math
\usepackage{amssymb,amsmath}
\usepackage{hyperref}

\newtheorem{theorem}{Theorem}[section]                          %
\newtheorem{lemma}[theorem]{Lemma}                              %
\DeclareMathAlphabet{\mathds}{U}{msb}{m}{n}
\newcommand{\R}{\ensuremath{\mathds {R}}}

\begin{document}
\title{Hard sphere crystallization gets rarer with increasing dimension}
\author{J.~A.~van~Meel}
\affiliation{FOM Institute for Atomic and Molecular Physics, Science Park 104, 1098 XG Amsterdam, The Netherlands}
\author{B.~Charbonneau}
\affiliation{Department of Mathematics, Duke University, Durham, North Carolina 27708, USA}
\author{A.~Fortini}
\affiliation{Theoretische Physik II, University of Bayreuth, Universit\"atsstrasse 30, Bayreuth, Germany}
\author{P.~Charbonneau}
\affiliation{Department of Chemistry, Duke University, Durham, North Carolina 27708, USA}
\date{\today}
\begin{abstract}
We recently found that crystallization of monodisperse hard spheres from the bulk fluid faces a much higher free energy barrier in four than in three dimensions at equivalent supersaturation, due to the increased geometrical frustration between the simplex-based fluid order and the crystal [J.~A.~van Meel, D.~Frenkel, and P.~Charbonneau, Phys. Rev. E \textbf{79} 030201(R) (2009)]. Here, we analyze the microscopic contributions to the fluid-crystal interfacial free energy to understand how the barrier to crystallization changes with dimension. We find the barrier to grow with dimension and we identify the role of polydispersity in preventing crystal formation. The increased fluid stability allows us to study the jamming behavior in four, five, and six dimensions and compare our observations with two recent theories [C.~Song, P.~Wang, and H.~A.~Makse, Nature \textbf{453} 629 (2008); G.~Parisi and F.~Zamponi, Rev. Mod. Phys, in press (2009)].
\end{abstract}
\maketitle

Structural glasses form under conditions where, though the thermodynamically stable state of the system is crystalline, the supersaturated fluid remains disordered. Instead of nucleating crystals, the fluid phase becomes steadily more viscous, until the microscopic relaxation processes become slower than the experimental or simulation timescales. A glass is then obtained. To avoid encountering a kinetic spinodal before falling out of equilibrium, good glass formers should therefore be poor crystallizers~\cite{kauzmann:1948}.
Geometrical frustration is one of the factors thought to slow the formation of ordered phases and thus assist glass formation~\cite{tarjus:2005}. Simple isotropic liquids are considered geometrically frustrated because the tetrahedron-based local order of the liquid cannot pack as a regular lattice. This scenario contrasts with what happens in a fluid of two-dimensional (2D) disks, where triangular order is locally as well as globally preferred and crystallization is particularly facile.

The initial formulation of geometrical frustration by Frank considered the optimal way for kissing spheres interacting via a Lennard-Jones model potential to cluster around a central sphere~\cite{frank:1952}. Frank found the icosahedral arrangement to be more energetically favorable than the cubic lattice unit cells. Though the original argument relies on the energetics of spurious surface effects~\cite{doye:1996}, mean-field solvation corrections leave the result unchanged~\cite{mossa:2003,mossa:2006}. For hard spheres, the icosahedron, which is the smallest maximum kissing-number Voronoi polyhedron, is also the optimally packed cluster. From the entropic standpoint the optimality of the structure therefore remains. More recently geometrical frustration has been couched in terms of the spatial curvature necessary to permit a defect-free lattice packing of tetrahedra (or, more generally, simplexes), which are the smallest building block in a Delaunay decomposition of space~\cite{nelson:2002,sadoc:1999}. This polytetrahedral scenario ascribes the presence of icosahedra to their singularly easy assembly from quasi-regular tetrahedra. Our recent study of 4D crystallization confirmed the simplex-based order in simple fluids as the source of frustration. The 4D optimal kissing cluster, which can also tile space in the densest known lattice packing, plays however no identifiable role in the liquid order~\cite{vanmeel:2009}. The observation that  optimal clusters, such as the icosahedron, are not singular is in agreement with the careful examination of the local fluid structure~\cite{anikeenko:2007,anikeenko:2008}, and offers a reasonable explanation for the limited amount of icosahedral order identified in experiments~\cite{schenk:2002,dicicco:2003,aste:2005,dicicco:2007} and simulations~\cite{kondo:1991,jakse:2003} of supercooled fluids.

Geometrical frustration also contributes to the nucleation barrier. In the absence of impurities or interfaces, crystallization proceeds through homogeneous nucleation in supersaturated solutions, as was spectacularly observed in container-less levitated metallic liquids~\cite{kelton:2003}. According to classical nucleation theory (CNT), homogeneous nucleation occurs through a rare fluctuation, whereby a crystallite that is sufficiently large for the bulk free energy gain to outweigh the interfacial free energy cost forms~\cite{volmer:1926}. Crystallization then spontaneously proceeds. The free energy difference between the ordered and disordered phases is fairly well understood microscopically in terms of the crystal packing efficiency. But the interfacial free energy contains a geometrical frustration contribution that is more challenging to interpret~\cite{nelson:1989}. Monodisperse hard spheres, the simplest system in which to study these effects, can indeed be supercooled in 3D, but they are notoriously bad glass formers. Our earlier study, which found crystallization barriers in 4D to be much higher than in equivalently supersaturated 3D fluids~\cite{vanmeel:2009}, suggest that 3D hard spheres are only moderately geometrically frustrated. In this article we improve on this qualitative assessment: to quantify geometrical frustration, we look at the dimensional trends and use bond-order parameters and the fluid-hard-wall interfacial free energy as a reference. Equipped with a microscopic understanding of geometrical frustration, we examine some of its consequences for 3D polydisperse spheres and consider how hard sphere crystallization evolves in higher dimensions. We find crystallization to become very rare, as was previously observed in systems of up to 6D~\cite{skoge:2006}. This rarity allows us to consider the consequences of the deeply supersaturated fluid branch on jamming, and compare the jamming results with the predictions of two recent theories~\cite{song:2008,parisi:2008}.

The plan of the paper is as follows. First, we complete the phase diagram of 5D and 6D hard spheres, and use quantitative tools to describe the fluid and crystal orders in the various dimensions. We then compute the fluid-hard-wall interfacial free energy in 2D, 3D and 4D, in order to quantify the contribution of geometrical frustration to the fluid-crystal interfacial free energy. Finally, we use these results to obtain the behavior of the free energy barrier to crystallization in higher dimensions.

\section{Methodology}
For convenience, in the rest of this article the particle diameter $\sigma$ sets the unit of length and the thermal energy $k_BT$ sets the unit of energy. For hard interactions this choice can be done without loss of generality, because entropy is the sole contributor to the free energy.

Spheres become less efficient at filling space with increasing dimension. Though with our choice of units the fluid densities of interest $\rho$ increase, the corresponding volume fraction $\eta$
\begin{equation}
\eta\equiv\rho V_d 2^{-d}=\rho\frac{\pi^{d/2}}{\Gamma(1+d/2) 2^d},
\label{eqn:eta}
\end{equation}
steadily decreases, because the volume of a $d$-dimensional hard sphere of radius $1/2$ in $\R^d$, $V_d 2^{-d}$, decreases faster than the crystal density increases. We report most quantities as volume fractions, but we also at times use $\rho$ if it simplifies the notation. %Also, for notational simplification, let $V_d\equiv V_d(1)$.

\subsection{Phase Diagram}
Because of computational limitations, 6D is the maximal dimension for which the phase diagram can reliably be obtained by simulation at this point. In a given dimension, in addition to the fluid phase we consider the crystal phase postulated to be the densest and a less dense crystal for reference. The densest known close-packed structures in 5D and 6D are degenerate through layering, the same way that hexagonal close-packed and face-centered cubic (fcc) crystals are degenerate through layering in 3D~\cite{conway:1995}. For convenience we choose the most symmetric of these as order phases, which are $D_5$ in 5D and $E_6$ in 6D respectively~\cite{conway:1988} (see Appendix~\ref{sect:genmatrix}). As in 3D, the impact of this decision on the phase diagram should be minimal~\cite{bolhuis:1997c}. With increasing dimensionality layered structures show a growing similarity in their local two- and three-particle distributions, because layering affects only one of a growing number of spatial dimensions. The choice of specific layered phase should thus have but a small impact on the structural analysis.

In order to precisely locate the freezing point, we compute the fluid and crystal hard sphere equations of state (EoS). Constant number of particles $N$, volume $V$, and temperature $T$ Monte Carlo (MC) simulations~\cite{frenkel:2002} give the radial pair distribution function $g(r)$, which once extrapolated at contact is related to the EoS
\begin{equation}
P/\rho= 1+B_2\rho g(1^+),
\end{equation}
where $P$ is the pressure and $B_2=V_d/2$ is the second virial coefficient~\cite{bishop:2005}. A sufficient number of MC cycles are used for the pressure to converge. Higher densities thus require longer simulations, because microscopic relaxation becomes sluggish. A minimum of 50,000 MC cycles are performed, but up to ten times that amount is used when needed. Starting configurations are obtained by slowly compressing the system and by equilibrating at each density of interest along the way.  For the fluid compressed beyond the freezing point, no crystallization is detected, which allows to thoroughly sample the metastable fluid, up until the microscopic relaxation becomes longer than the simulation time. The EoS are obtained for systems
of $2048$ ($D_4$), $4096$ (4D fluids and $A_4$), $3888$ (5D fluids and $D_5$), $14400$ ($A_5$), $2048$ ($D_6$), $10000$
(6D fluids), and $17496$ ($E_6$) particles. To locate the fluid-crystal coexistence regime, we determine the absolute Helmholtz free energy per particle $f$ of the crystal using the Einstein-crystal method~\cite{frenkel:1984} at a reference point: for $D_4$ and $A_4$ crystals we use $\eta=0.37$, for $D_5$ and $A_5$ $\eta=0.21$, and for $D_6$ and $E_6$  $\eta=0.12$. The excess free energy at other crystal densities is then obtained by thermodynamic integration of the EoS~\cite{frenkel:2002}. To obtain the fluid free energy the EoS is integrated from the ideal gas limit. The chemical potential
\begin{equation}
\mu(\rho)=f+P/\rho
\end{equation}
gives the fluid-crystal coexistence pressure $P^{\mathrm{coex}}$ and $\mu^{\mathrm{coex}}$ by finding where it is equal for both phases  $\Delta \mu(P)=0$. The densities of the coexisting phases is then obtained from constant $N$, $P^{\mathrm{coex}}$, and $T$ MC simulations. This approach is formally equivalent to the common tangent construction, but we find it to be numerically more efficient.

\subsection{Order Parameters}
\label{sec:metho-order}
To characterize the structure of the fluid and crystal phases we need a criterion to quantify local ordering. Studies in 2D and 3D suggest that order parameters derived from rotationally-invariant combinations of the $m$ different spherical harmonics $Y_l^m$ of degree $l$ might suffice~\cite{nelson:1981,tenwolde:1996,auer:2001}.
Here, we consider second- and third-order invariants, which are sensitive to the degree of spatial orientational correlation of the vectors that join neighboring particles. For a proper choice of $l$ the (renormalized) invariant's (absolute) value is one for a perfect crystal and close to zero for a perfectly isotropic fluid.
%----------
Though a 4D canonical spherical harmonics basis~\citep[Sec.~9.6]{hamermesh:1962} and both its second- and third-order invariants~\cite{nelson:1984} are known, in higher dimensions it rapidly becomes analytically intractable to identify a basis composed of weight vectors for the representation of SO($d$)~\cite{rowe:2004}. It is therefore more convenient to rewrite the invariants as polynomials of the vector inner products. For the second-order invariants, one simply uses the Gegenbauer polynomials $G_l^{d/2-1}$ obtained from the sum rule~\cite[Thm.~9.6.3]{SpecialFunctions}. For instance, the sum over the $(l+1)^2$ 4D spherical harmonics for unit vectors $\hat{\mathbf{r}}_i$ can be rewritten as
\begin{equation}
G_l^1(\hat{\mathbf{r}}_1\cdot\hat{\mathbf{r}}_2)=\frac{2\pi^2}{(l+1)^2}
\sum_{m=1}^{(l+1)^2}Y_l^m (\hat{\mathbf{r}}_1) \overline{Y_l^{m}
(\hat{\mathbf{r}}_2)}.
\end{equation}
The second-order local bond-order correlator $q_l(i,j)$ is then obtained by summing over the $N(i)$ and $N(j)$ nearest neighbors of particles $i$ and $j$ conveniently chosen to be within a distance equal to the first minimum of $g(r)$. By letting the indices $\alpha$ and $\beta$ run over these neighbors we find~\cite{vanmeel:2009}
\begin{equation}
q_l(i,j)=\mathbf{q}_l(i)\cdot\mathbf{q}_l(j)=\frac{
\sum_{\alpha=1}^{N(i)}\sum_{\beta=1}^{N(j)}G_l^{d/2-1}
(\hat{\mathbf{r}}_{i \alpha}\cdot\hat{\mathbf{r}}_{j \beta})}{N(i)N(j)}.
\label{eq:ord_param}
\end{equation}
%{\bf [KvM]: The two-particle second-order invariant $q_l(i,j) = \mathbf{q}_l(i) \cdot \mathbf{q}_l^*(j)$, rather than a single-particle $q_l(i) = \mathbf{q}_l(i) \cdot \mathbf{q}_l^*(i) = q_l(i,i)$ or global second-order invariant $Q_l = [ 4\pi / (2l+1) \sum_{m=-l}^{l} |\sum_{i=1}^N \sum_{j=1}^{N(i)}Y_l^m(\hat{\mathbf{r}}_{ij}) / \sum_{i=1}^{N} N(i) ]^{1/2}$, is known as a reliable basis for the identification of individual crystalline particles in 3D~\cite{tenwolde:1996} and 4D~\cite{vanmeel_2009}, and allows for a more sensitive analysis of the fluid structure.}
This two-particle second-order invariant $q_l(i,j)$, rather than a single-particle $q_l(i,i)$ or a global second-order invariant $Q_l= \left[ \sum_i \sum_j N(i) N(j) q_l(i,j) \right]^{1/2} / \sum_i N(i)$, is known to be a reliable basis for the identification of individual crystalline particles in 3D~\cite{tenwolde:1996} and 4D~\cite{vanmeel:2009}, and allows for a more sensitive analysis of the fluid structure.

We now develop an approach to obtain third-order rotationally invariant polynomials $\tilde{w}_l$ analogous to the Gegenbauer polynomials. This approach, like the one described above for the second-order invariant, has the important advantage that we do not need a prior knowledge of a canonical spherical harmonics basis. A classical theorem due to Weyl says that any polynomial in $m$ sets of variables $X_1,\ldots, X_m\in \R^d$  invariant under the diagonal action \[g\cdot f(X_1,\ldots, X_m)=f(gX_1,\ldots,gX_m)\text{ for } g\in\mathrm{SO}(d)\]
can be expressed in terms of the inner products $\langle X_i,X_j\rangle $ and the determinants $\det[{X_{i_1} \ldots X_{i_d}}]$.
For third-order invariants ($m=3$) in $d\geq 4$, all the determinants are zero, and we are able to write the invariant
polynomial  in $X=(X_1,\ldots,X_d)$, $Y=(Y_1,\ldots,Y_d)$,
and $Z=(Z_1,\ldots, Z_d)$ in terms of the various inner products $x=\langle
X,X\rangle,y=\langle Y,Y\rangle ,z=\langle Z,Z\rangle$ and
$a=\langle X,Y\rangle,b=\langle X,Z\rangle ,c=\langle Y,Z\rangle$.

Let $f$ be a polynomial in $X,Y,Z$. Suppose that $f$ is invariant under the diagonal action of $\mathrm{SO}(d)$. Then there is a polynomial $p(x,y,z,a,b,c)$ such that \[f(X,Y,Z)=p(x,y,z,a,b,c).\]

\begin{lemma}Suppose $f$ is homogeneous of degree $l$ in $X$, $Y$, and $Z$ separately, and is therefore homogeneous  of degree $3l$ overall. Then
\[p(\lambda^2x,\mu^2y,\nu^2z,\lambda\mu a,
\lambda\nu b,\mu\nu c)=(\lambda\mu\nu)^lp(x,y,z,
a , b,c).\]
In particular, $p$ is homogeneous of degree $3l/2$ overall.
\label{lemma:hom}
\end{lemma}

Let
\begin{widetext}
\begin{align*}
D_X(p)&\equiv
2d{\frac{\partial p}{\partial x}}+ 4x{\frac{\partial^{2}p}{\partial{x}^{2}}}+ 4a{\frac{\partial^{2}p}{\partial a\partial x}}+ 4b{\frac{\partial^{2}p}{\partial b\partial x}}+ 2c{\frac{\partial^{2}p}{\partial a\partial b}}+ y{\frac{\partial^{2}p}{\partial{a}^{2}}}+ z{\frac{\partial^{2}p}{\partial{b}^{2}}},\\
D_Y(p)&\equiv
2d{\frac{\partial p}{\partial
y}}+4y{\frac{\partial^{2}p}{\partial{y}^{2}}}+4a{\frac {\partial^{2}p}
{\partial a\partial y}}+4c{\frac {\partial^{2}p}{\partial
c\partial y}}+2b{\frac {\partial^{2}p}{\partial a\partial c}}+x{
\frac {\partial^{2}p}{\partial{a}^{2}}}+z{\frac
{\partial^{2}p}{\partial{c}^{2}}},\\
D_Z(p)&\equiv 2d{\frac{\partial p}{\partial z}}+ 4z{\frac{\partial^{2}p}{\partial{z}^{2}}}+ 4b{\frac{\partial^{2}p}{\partial b\partial z}}+ 4c{\frac{\partial^{2}p}{\partial c\partial z}}+ 2a{\frac{\partial^{2}p}{\partial b\partial c}}+ x{\frac{\partial^{2}p}{\partial{b}^{2}}}+ y{\frac{\partial^{2}p}{\partial{c}^{2}}}.\\
\end{align*}
\end{widetext}

\begin{lemma}The operators above satisfy
\[\nabla^2_X p(x,y,z,a,b,c)=(D_Xp)(x,y,z,a,b,c),\]
and similarly for $Y$ and $Z$.
In particular if $f$ is harmonic in $X$, $Y$ and $Z$, then
$D_X(p)=D_Y(p)=D_Z(p)$=0.
\label{lemma:Laplacian}
\end{lemma}

The proof is an exercise in using the chain rule.

Using Lemmas \ref{lemma:hom} and \ref{lemma:Laplacian}, we set up a system of equations for the coefficients of the polynomial $p(x,y,z,a,b,c)$ corresponding to a $\mathrm{SO}(d)$ invariant polynomial $f(X,Y,Z)$ of degree $l$ and harmonic in each of $X$, $Y$, and $Z$ separately.  We can solve this system of equation, and once we choose the normalization, say $p(1,1,1,0,0,0)=1$, there is a unique solution. Setting $x=y=z=1$ restricts the obtained function on the unit sphere. We call the resulting function $\tilde w_l^d(a,b,c)$. Examples are given in Appendix~\ref{sect:thirdorder}. For the reader cognisant of representation theory, note that $w_l^d(X,Y,Z)=\tilde w_l^d(a,b,c)$ is a basis for the copy of the irreducible representation in the triple tensor product $H_l(S^{d-1})\otimes H_l(S^{d-1})
\otimes H_l(S^{d-1})$.

As in the second-order case, those polynomials allow us to rewrite the third-order local bond-order correlator $W_l(i)$ up to a dimension-dependent multiplicative constant $c^d_l$
\begin{equation}
W_l(i)=c^d_{l}\frac{\sum_{\alpha,\beta,\delta}^{N(i)}\tilde{w}^d_l(\hat{\mathbf{
r}}_{
i\alpha
}\cdot\hat{\mathbf{r}}_{i\beta},\hat{\mathbf{r}}_{i\alpha}\cdot
\hat{\mathbf{r}}_{i\delta},\hat{\mathbf{r}}_{i\beta}\cdot\hat{\mathbf{r}}_{
i\delta})}{{2^{d-2} [N(i)]^3 [q_l(i,i)]^{3/2}}}.
\end{equation}
In 3D and 4D, the constant
$c^d_l$ can be set by comparing with the expression available in the literature.
In higher dimensions, we choose the normalization for which the polynomial
equals unity when evaluated on three orthogonal unit vectors i.e.,
$c^d_l \tilde{w}^d_l(0,0,0)=1$. Note that because of the rotational symmetry,
triplets with
permuted indices can be summed only once by correcting for the
multiplicity. This simplification offers an important computational advantage.
Though the use of rotationally-invariant polynomials for the computation of the
bond-order parameters is mainly used for analytical convenience, it is also worth
noting that for large $l$ and at low densities, it is computationally
more efficient than the standard spherical harmonics decomposition, and that their algebraic simplicity minimizes the risks of error at the implementation stage.

\subsection{Wall Cleaving Surface Tension}
The 2D, 3D, and 4D hard sphere fluid-hard-wall interfacial free energy $\gamma_{\mathrm{f-w}}$ is calculated through MC simulation using the higher-dimensional generalization of an earlier thermodynamic integration scheme~\cite{fortini:2005}. We start from a system periodically confined by both sides of a hard wall perpendicular to the $x$ axis and make the wall gradually more penetrable until the bulk system is obtained. Confinement is achieved by introducing the auxiliary Hamiltonian
\begin{equation}
H_{\lambda}= \sum_{i,j}^{N} V_{\mathrm{HS}}(r_{ij})  +\lambda \sum_{i=1}^{N} V_{\mathrm{w}}(x_{i}),
 \label{E:ham}
\end{equation}
where $V_{\mathrm{HS}}$ is the hard-core exclusion between hard spheres and $V_{\mathrm{w}}$ is the penetrable wall potential
\begin{equation}
V_{\mathrm{w}}(x)= \left \{ \begin{array}{ll}
 \exp(-2 x) & \textrm{ if $|x| < \sigma/2$ } \\
0 & \textrm{ otherwise}
\end{array} \right . \
\end{equation}
for a sphere a distance $x$ from the wall.
This truncated exponential function is known to provide a high numerical accuracy route for $\gamma_{\mathrm{f-w}}$ computation~\cite{fortini:2007,fortini:2005}.
When the coupling parameter $\lambda=\infty$, hard spheres are confined by a hard wall, while for
$\lambda=0$ the bulk fluid limit is recovered.
The interfacial free energy is obtained by Kirkwood integration
\begin{equation}
\gamma_{\mathrm{f-w}}=\frac{1}{2A}\int_{0}^{\infty} d \lambda \left\langle  \frac{\partial H_{\lambda}}{\partial \lambda} \right\rangle_{\lambda},
\end{equation}
where $A$ the area of a single side of the wall. In practice the integral is solved using a 21 point Gaussian-Kronrod formula in a finite interval $\lambda \in (0,\lambda_{\rm max})$ with $\lambda_{\mathrm{max}}$ chosen to approximate arbitrarily closely a hard wall.

\subsection{Generalized Classical Nucleation Theory}\label{sec:generalized_cnt}
Classical  nucleation theory (CNT)~\cite{volmer:1926} considers contributions from chemical potential difference between the bulk phases
and the fluid-crystal interfacial free energy $\gamma_{\mathrm{f-x}}$ of a spherical crystallite to obtain a free energy functional
\begin{equation}
\Delta G(n)=A_d (n/\rho_{\mathrm{x}})^{(d-1)/d} \gamma_{\mathrm{f-x}} -n\Delta\mu,
\label{eq:cnt}
\end{equation}
of the number of particles $n$ in the crystallite. The functional further depends on the crystal density $\rho_{\mathrm{x}}$ at the supersaturated fluid pressure and on a geometrical prefactor $A_d$. For hard spheres $A_d=S_{d-1} V_d^{1/d-1}$, where $S_{d-1}=d V_d$\label{Sd-first-occur} is the surface area of a $d$-dimensional
unit sphere. The resulting barrier height at the critical cluster size $n^*$ is
\begin{equation}
\Delta G^*(n^*)=\frac{ (d-1)^{d-1} \pi^{d/2}}{\Gamma(d/2+1)2^d} \, \frac{\gamma_{\mathrm{f-x}}^d}{\rho_{\mathrm{x}}^{d-1}\Delta\mu^{d-1}},
\label{eq:barrheight}
\end{equation}
and in the high dimensional limit the barrier asymptotically approaches
\begin{equation}
\Delta G^*(n^*)\sim(2\pi e d)^{d/2}\frac{\gamma_{\mathrm{f-x}}^d}{\rho_{\mathrm{x}}^{d-1}\Delta\mu^{d-1}}.
\label{eq:barrtrend}
\end{equation}

The rate of nucleation per unit volume $k$ is
then $k=\kappa \exp(-\Delta G^*)$, where $\kappa$ is a kinetic prefactor proportional to the diffusion
coefficient in the fluid phase~\cite{auer:2001}. The kinetic prefactor has a weak dimensionality dependence that we won't consider here. Though schematic, this level of theory is sufficient to clarify
the contribution of geometrical frustration to the crystallization barrier through an analysis of $\gamma_{\mathrm{f-x}}$. Within the CNT framework
geometrical frustration between ordered and disordered phases should lead to a relatively large $\gamma_{\mathrm{f-x}}$, and thus
to a high crystallization free energy barrier. On the contrary, geometrically similar phases should have small $\gamma_{\mathrm{f-x}}$ and $\Delta G^*(n^*)$.

\section{Results and Discussion}

\subsection{Phase Diagram and Jamming}

\begin{figure*}
\center{
\includegraphics[width=\columnwidth]{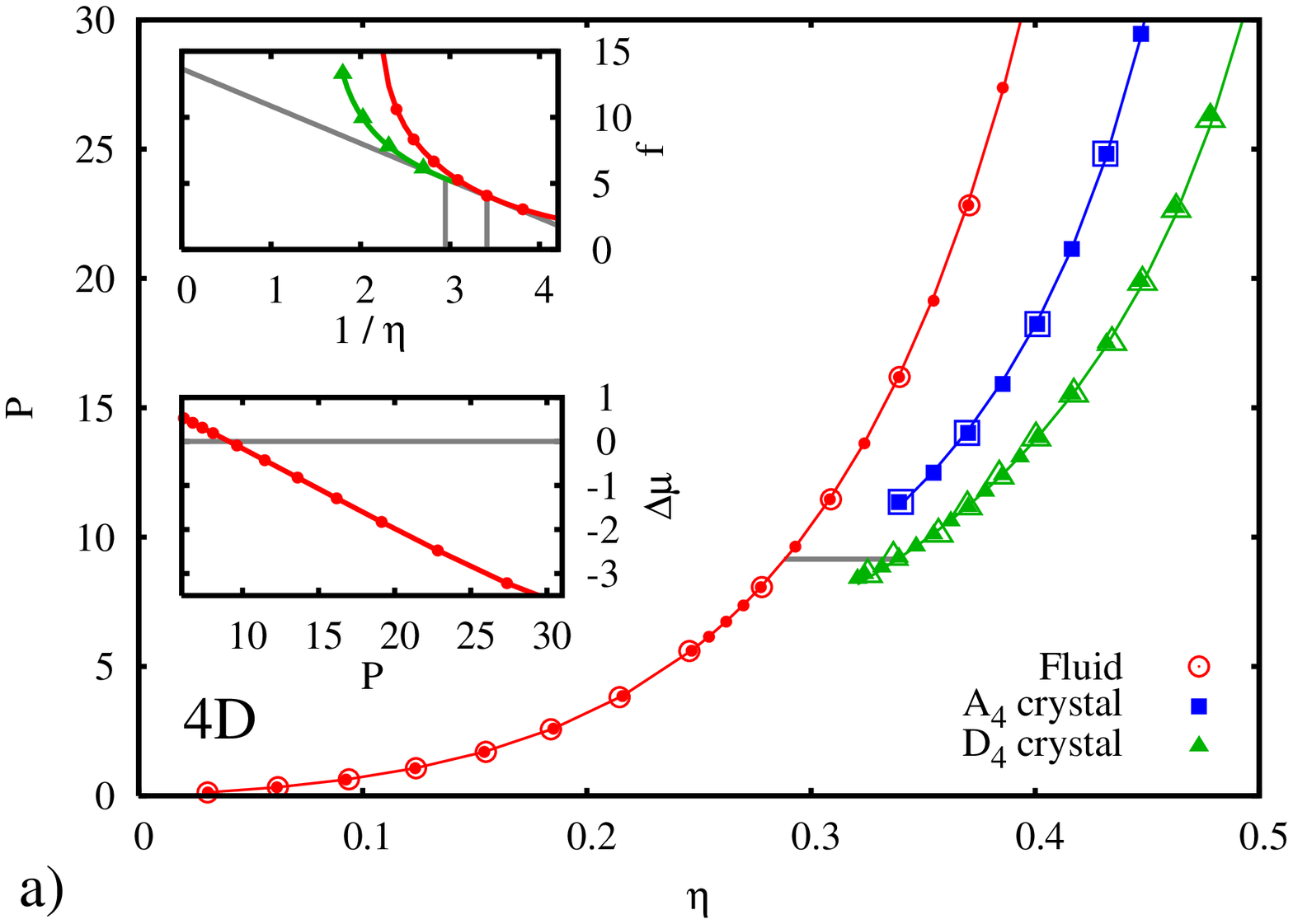}\hfill
\includegraphics[width=\columnwidth]{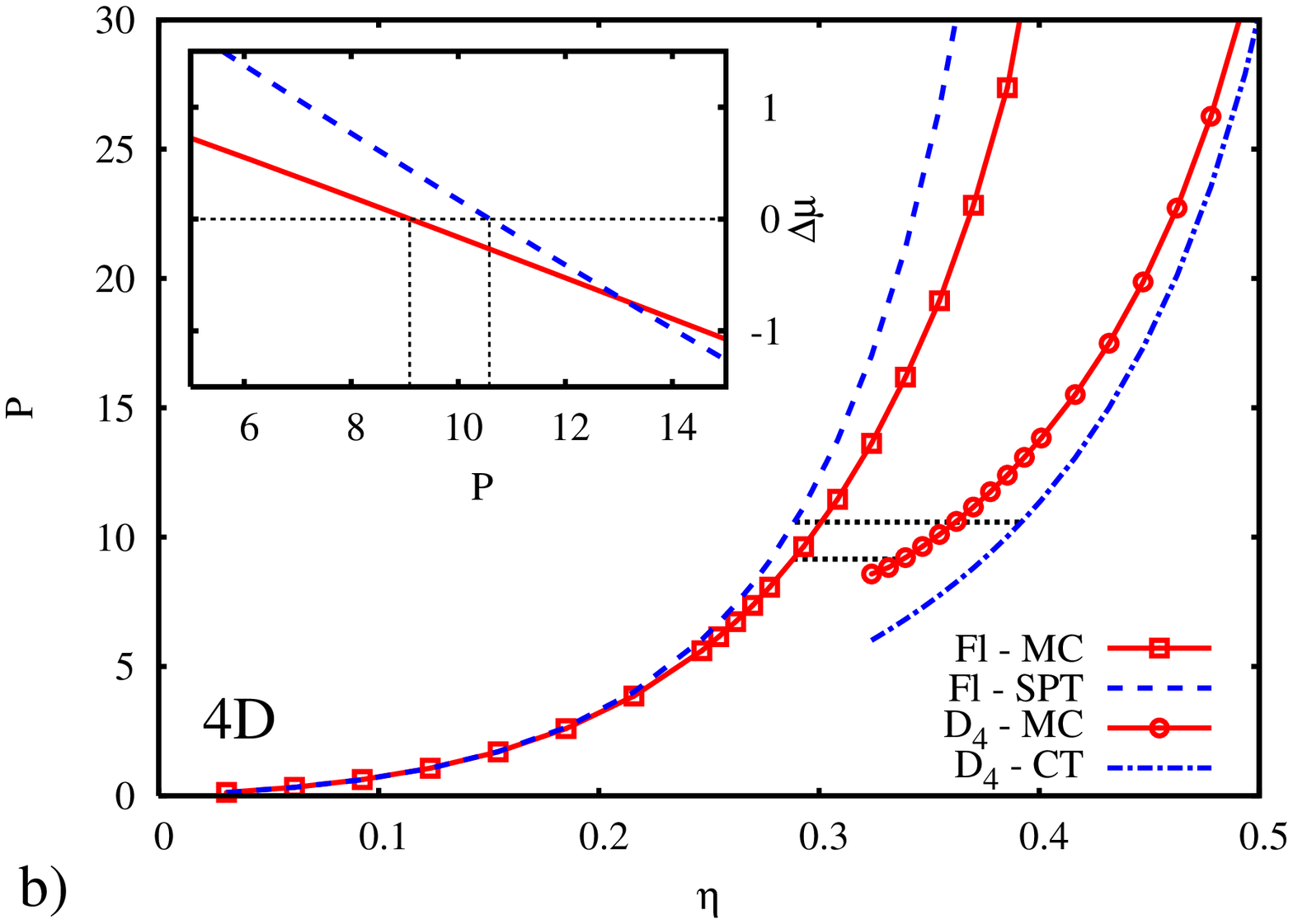}}\\
\center{\includegraphics[width=\columnwidth]{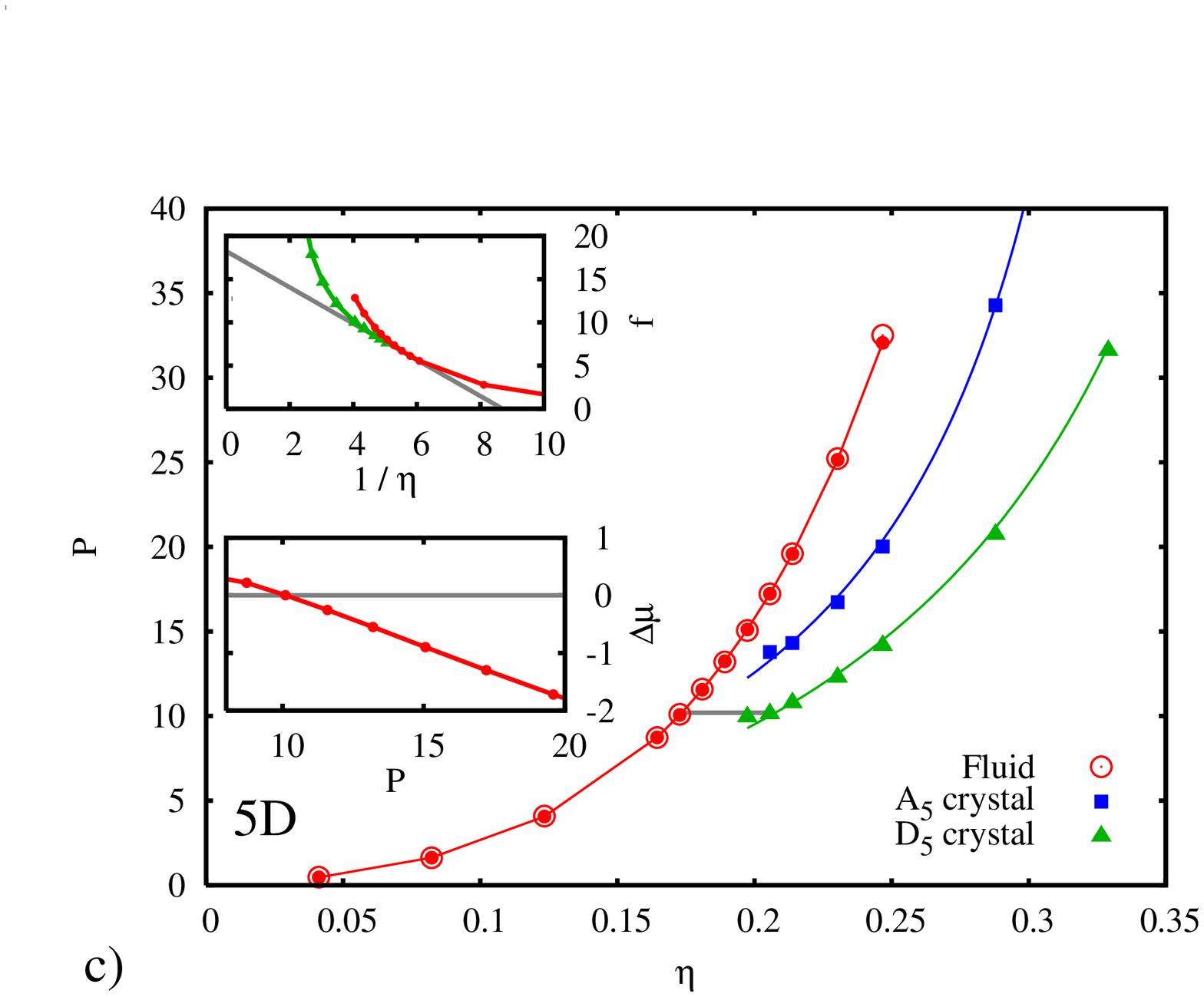}\hfill
\includegraphics[width=\columnwidth]{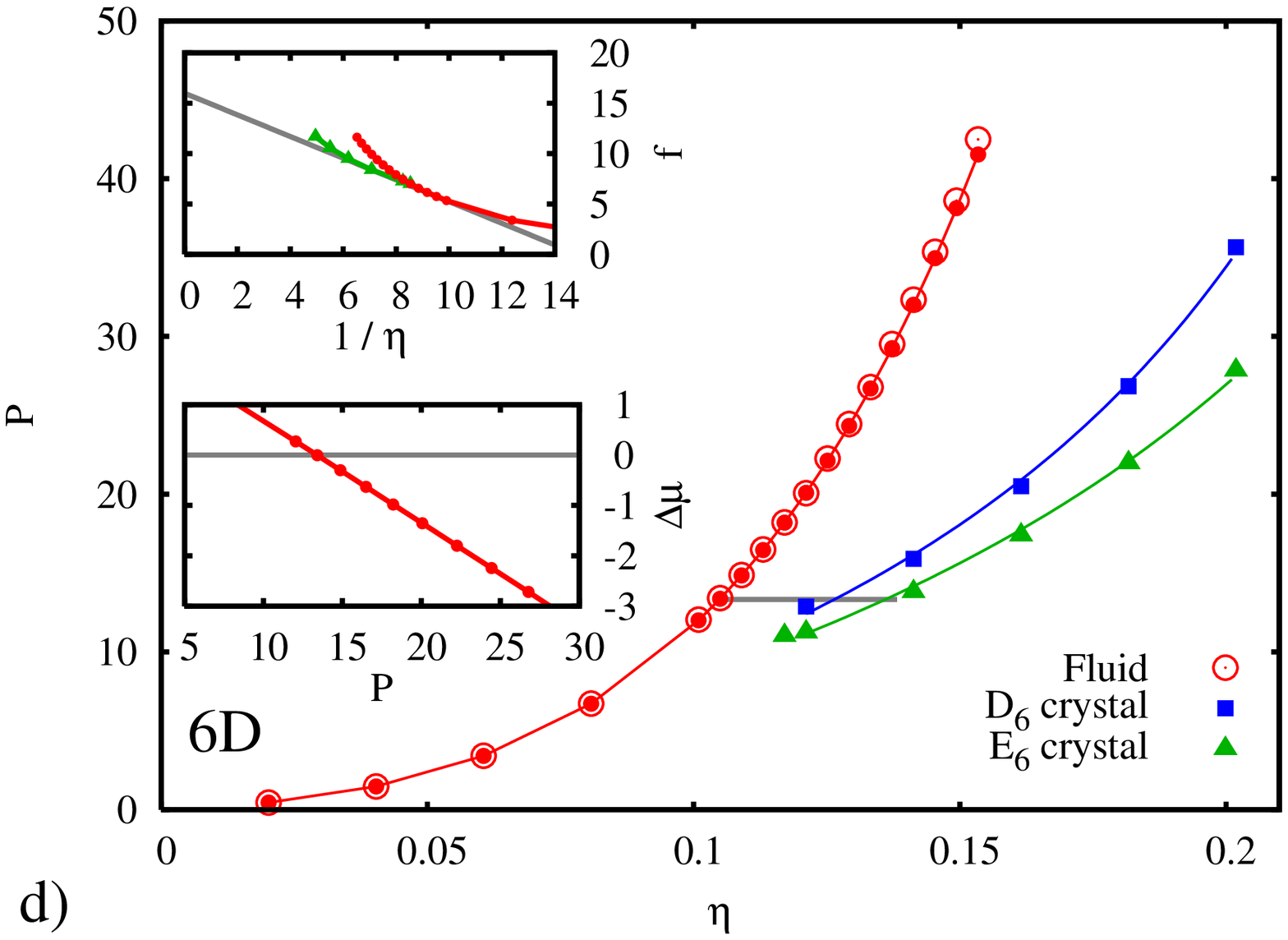}}
\caption[Phase diagram]{(Color online) Monte Carlo EoS of the fluid and the two densest known ordered phases in 4D~\cite{vanmeel:2009} (a), 5D (c), and 6D (d) computed  at constant V (filled symbols) and constant P (open symbols), with Pad\'e approximants of the virial expansion for the fluid~\cite{bishop:2005}, and the Speedy fits to the crystal phase results~\cite{speedy:1998} (solid lines). Insets give $\Delta \mu$ and the common tangent construction for determining coexistence between the fluid and the densest crystal phase. The additional panel for 4D (b) contrasts MC and SPT/CT EoS as well as the resulting coexistence determination (inset).}
\label{fig:phasediagram}
\end{figure*}

\begin{table*}
\begin{ruledtabular}
\begin{tabular}{|c|c|c|c|c|c|c|}
  $d$ & $P^{\mathrm{coex}}$ &$\mu^{\mathrm{coex}}$  & $\eta_{\mathrm{f}}-\eta_{\mathrm{x}}$  & Modified Ref.~\cite{skoge:2006} $\eta_{\mathrm{f}}-\eta_{\mathrm{x}}$  & Corrected SPT/CT~\cite{finken:2001} $\eta_{\mathrm{f}}-\eta_{\mathrm{x}}$ & $\eta_{\mathrm{cp}}$\\
    \hline
  3~\cite{hoover:1968} & 11.564 & 17.071 & $0.494-0.545$ & - &- & 0.741 \\
  4~\cite{vanmeel:2009} & 9.15 & 13.7  & $0.288-0.337$ & $0.29-0.35$ & $0.29-0.39$ & 0.617\\
  5 & 10.2 & 14.6  & $0.174-0.206$ & $0.18-0.22$ & $0.17-0.24$ & 0.465\\
  6 & 13.3 & 16.0  & $0.105-0.138$ & - & $0.10-0.16$ & 0.373\\
\end{tabular}
\end{ruledtabular}
\caption[Coexistence parameters]{Coexistence parameters from Monte Carlo simulations compared with previous simulation estimates (see text) and the corrected SPT/CT results (see text). The volume fraction of the densest known lattice $\eta_{\mathrm{cp}}$ is also included for reference~\cite[Chap 1. $\S$ 1.5]{conway:1988}} \label{tab:coexistence}
\end{table*}

\begin{table}
\begin{ruledtabular}
\begin{tabular}{|c|c|c|c|c|}
  $d$ &
  $\eta^{\mathrm{fV}}_{\mathrm{MRJ}}$ & $\eta^{\mathrm{comp}}_{\mathrm{MRJ}}$~\cite{skoge:2006} & $\eta^{\mathrm{stat}}_{\mathrm{MRJ}}$ & $\eta^{\mathrm{rep}}_{\mathrm{MRJ}}$~\cite{parisi:2008}\\
    \hline
  4 & 0.47 & 0.46  & 0.43 & 0.49 \\
  5 & 0.31 & 0.31  & 0.28 & 0.33 \\
  6 & 0.21 & 0.20  & 0.17 & 0.22 \\
\end{tabular}
\end{ruledtabular}
\caption[MRJ]{Volume fraction of the maximally-random jammed states obtained from free-volume extrapolation $\eta^{\mathrm{fV}}_{\mathrm{MRJ}}$, direct compression~\cite{skoge:2006} $\eta^{\mathrm{comp}}_{\mathrm{MRJ}}$, an extension of the statistical mean-field approach of Ref.~\cite{song:2008} $\eta^{\mathrm{stat}}_{\mathrm{MRJ}}$, and the replica method mean-field approach~\cite{parisi:2008} $\eta^{\mathrm{rep}}_{\mathrm{MRJ}}$} \label{tab:MRJ}
\end{table}

The computed fluid EoS agrees with earlier 4D~\cite{vanmeel:2009}, 5D~\cite{michels:1984,luban:1990,bishop:2005,lue:2005}, and 6D~\cite{lue:2006} simulation results as well as the 5-4 Pad\'e approximants of the virial expansion~\cite{bishop:2005,bishop:2007} (Fig.~\ref{fig:phasediagram}).  Small deviations are only observed at the highest densities, where the expansion is less accurate~\cite{bishop:2005,lue:2006}. Crystal phase EoS for the $D_4$ and $D_5$ lattice geometries in 4D and 5D respectively were first obtained from simulation in the early 80's, but without reference free energies~\cite{michels:1984,skoge:2006}, and we are not aware of any 6D simulation results. As expected from free volume arguments, the densest known lattice is the phase with the lowest pressure at densities where it is mechanically stable, and is the most free energetically favorable of the ordered phases. Assuming that the crystallization kinetics is controlled by the free energy barrier height, the most stable ordered phase should be the only relevant one of hard spheres. The generation of crystallites with other symmetries is only possible at much higher pressures and with a smaller thermodynamic drive. The scaling of the fluid-crystal interfacial free energy with pressure energy suggests that neither phase would have a significant advantage over the other from that respect (cf. Sect.~\ref{sect:gammaf-x}).

The fluid-crystal coexistence conditions from simulation for hard spheres in 5D and 6D, along with the earlier 3D~\cite{hoover:1968} and 4D~\cite{vanmeel:2009} results are reported  in Table~\ref{tab:coexistence}. Skoge et al. offered upper bounds to the 4D and 5D coexistence regimes by using the pressure of the fluid at the density at which the simulated $D_d$ crystal becomes mechanically unstable as an estimate of coexistence pressure $P^{\mathrm{coex}}$~\cite{skoge:2006}. A more accurate estimate of $P^{\mathrm{coex}}$ can be obtained from the same data by using instead a quasi-Maxwell construction around the limit of mechanical stability~\cite{streett:1974}. We include the results of this last analysis and the coupled fluid scaled-particle theory (SPT) and crystal cell theory (CT) coexistence determination of Finken et al.~\cite{finken:2001} in Table~\ref{tab:coexistence} as well. To the best of our knowledge, density functional theory (DFT) coexistence parameters have only been reported for the non-equilibrium 4D fluid-$A_4$ pair of phases~\cite{colot:1986}, which does not lend itself to a meaningful comparison with simulations. Finken et al. do refer to DFT coexistence calculations, but do not report their results for 4D to 6D~\cite{finken:2001}.

Our MC results are at least an order or magnitude more precise than the estimates from Ref.~\cite{skoge:2006}, which permits a clearer assessment of the SPT/CT predictions. The SPT/CT analysis correctly captures certain dimensional trends. For instance, the relative difference in crystal and fluid volume fraction at coexistence $\Delta \eta^{\mathrm{coex}}/\eta_{\mathrm{x}}$, which is thought to go to unity for large dimensions~\cite{finken:2001}, does increase appreciably from below $10\%$ in 3D to over $20\%$ in 6D. And the crystal volume fraction at coexistence $\eta_{\mathrm{x}}$ decreases relatively faster than the close-packed volume fraction $\eta_{\mathrm{cp}}$, which leaves the phase diagram dominated, in percentage of the accessible densities, by the ordered phase~\cite{footnote:3}. Finken et al., however, incorrectly implement the common tangent construction, which explains why their reported coexistence regimes~\cite{finken:2001} are well above our simulation results. Once corrected, the theoretical SPT/CT predictions more closely follow the simulation results (see Table~\ref{tab:coexistence}). SPT shows a fairly good agreement with the fluid EoS, and, as expected from the third-order virial coefficient, overshoots the fluid pressure at high densities~\cite{finken:2001} (see Fig.~\ref{fig:phasediagram}). CT however significantly underestimates the crystal pressure near coexistence and the effect does not go away with dimension. The high compressibility of hard sphere crystals near the limit of mechanical stability is a collective effect that is not captured by the mean-field nature of the theory. The cancelation of errors leaves SPT/CT coexistence densities reasonably on target, but SPT/CT overestimates the width of the coexistence densities and the coexistence pressure. It is unclear if this effect vanishes with dimension.

It is interesting to note that $P^{\mathrm{coex}}$ does not change monotonically with dimension, but goes through a minimum in 4D. The nonmonotonic behavior of $P^{\mathrm{coex}}$ might be due to $D_4$'s particularly well-suited nature to fill 4D Euclidian space. A $D_4$ lattice can be generated by placing a sphere in each of the voids of a 4D simple cubic lattice. These new spheres are equidistant to the ones on the simple cubic frame, so the resulting lattice is twice denser. The corresponding construction $D_3^*$, or body centered-cubic lattice, in 3D packs much less efficiently, because the simple cubic frame needs to be extended to insert the new spheres. Though $D_4$ does not appear singular in the dimensional trend of dense packings~\cite[Chap. 1, $\S1.5$]{conway:1988}, its specificity might simply be overshadowed by other dimensional trends to which $P^{\mathrm{coex}}$ is less sensitive. We therefore conjecture that the non-monotonic coexistence pressure is a symmetry signature that should also be present in 8D, 12D, 16D, and 24D, where other singularly dense lattices are known to exist.

Another noteworthy observation concerns the high pressure limit of the fluid EoS. The fluid results presented in this article are only for systems in equilibrium or in metastable equilibrium, i.e. the initial configurations are prepared in the fluid state and the simulations are run much longer than the fluid microscopic relaxation timescale, though much shorter than the nucleation timescale. But because no crystallization takes place on the simulation time scale, only the growing microscopic relaxation timescale limits the range of densities can be reached in simulations. The EoS can thus obtained for supersaturated systems at much higher pressures than in 3D (see Sect.~\ref{sect:CNT_discuss}). We extract the infinite-pressure compression limit of the densest supersaturated fluid point on the EoS from the free-volume functional form for pressure $P^{\mathrm{fV}}$ suggested in Ref.~\cite{kamien:2007}
\begin{equation}
P^{\mathrm{fV}}(\eta)=\frac{\eta}{V_d 2^{d}\left[1-\left(\eta/\eta^{\mathrm{fV}}\right)^{1/d}\right]}.
\label{eq:freeV}
\end{equation}
This functional form, which is supported by the analysis of Parisi and Zamponi~\cite{parisi:2008} for 4D fluids, corresponds to a non-equilibrium compression so rapid that no microscopic relaxation can take place. Because microscopic relaxation becomes increasingly slow large at high fluid densities, it approximates the compression algorithm of Ref.~\cite{skoge:2006}. The volume fraction of the disordered jammed system corresponding to the densest equilibrated fluid is obtained by solving for $\eta^{\mathrm{fV}}$. The parameter $\eta^{\mathrm{fV}}$ is not the volume fraction of the maximally-dense random jammed (MRJ) state per se. But were it possible for the fluid to avoid crystallization indefinitely, which is obviously an unphysical but useful abstraction, it should asymptotically approach it from below. A quenched  compression started from the non-crystallizing, but otherwise equilibrated fluid branch at higher supersaturations, where microscopic relaxation is even slower, would give a denser $\eta^{\mathrm{fV}}$~\cite{parisi:2008}. The high densities achieved here should thus give a fairly reliable lower bound to $\eta_{\mathrm{MRJ}}$. The resulting values are in very good agreement with the volume fractions obtained by direct non-equilibrium compression~\cite{skoge:2006} (see Table~\ref{tab:MRJ}). A recent statistical mean-field theory of jamming that shows a surprisingly high accuracy with the experimental and simulation $\eta_{\mathrm{MRJ}}$ in 3D~\cite{song:2008}, would be expected to perform similarly well, if not better, when dimensionality is increased, due to the growing number of nearest neighbors. On reproducing the arguments of Ref.~\cite{song:2008} for higher dimensions (see Appendix~\ref{sect:statjam}), we find that though the accuracy is still quite good, it is not as high as for 3D. The propagated relative error of the statistical mean-field treatment is about $5$--$10\%$, which suggests that the high accuracy of the analytical 3D results of Ref.~\cite{song:2008} might be fortuitous or that it is particularly sensitive to the choice of scaling assumptions. The mean-field treatment based on the replica method offers however predictions that more consistent with the simulation results~\cite{parisi:2008}. The dimensional trends are more similar, and the simulation results are slightly smaller than the theoretical predictions, which is precisely where they are expected to be~\cite{parisi:2008}.

\subsection{Bond-Order Correlators}
\label{sect:bond-orderdist}
\begin{figure}
\center
\includegraphics[width=\columnwidth]{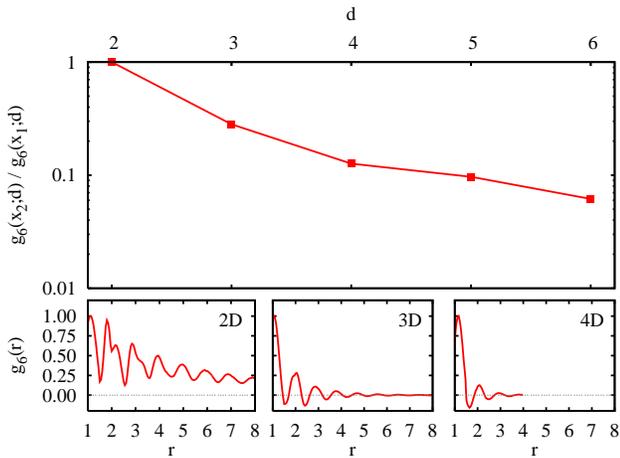}
\caption[$g_6(r)$]{(Color online) Impact of dimensionality on the decay of orientational order. Top: Second to first peak ratio of the radial decay of the orientational order parameter $g_6(r)$ at fluid coexistence. The line is a guide to the eye. Bottom: Decay of $g_6(r)$ in the fluid near the hexatic phase in 2D and at coexistence in 3D and 4D. The 5D and 6D plots (not shown) are qualitatively similar to the 4D results.}
\label{fig:bondcorrelator}
\end{figure}

\begin{figure*}
\center{\includegraphics[width=\columnwidth]{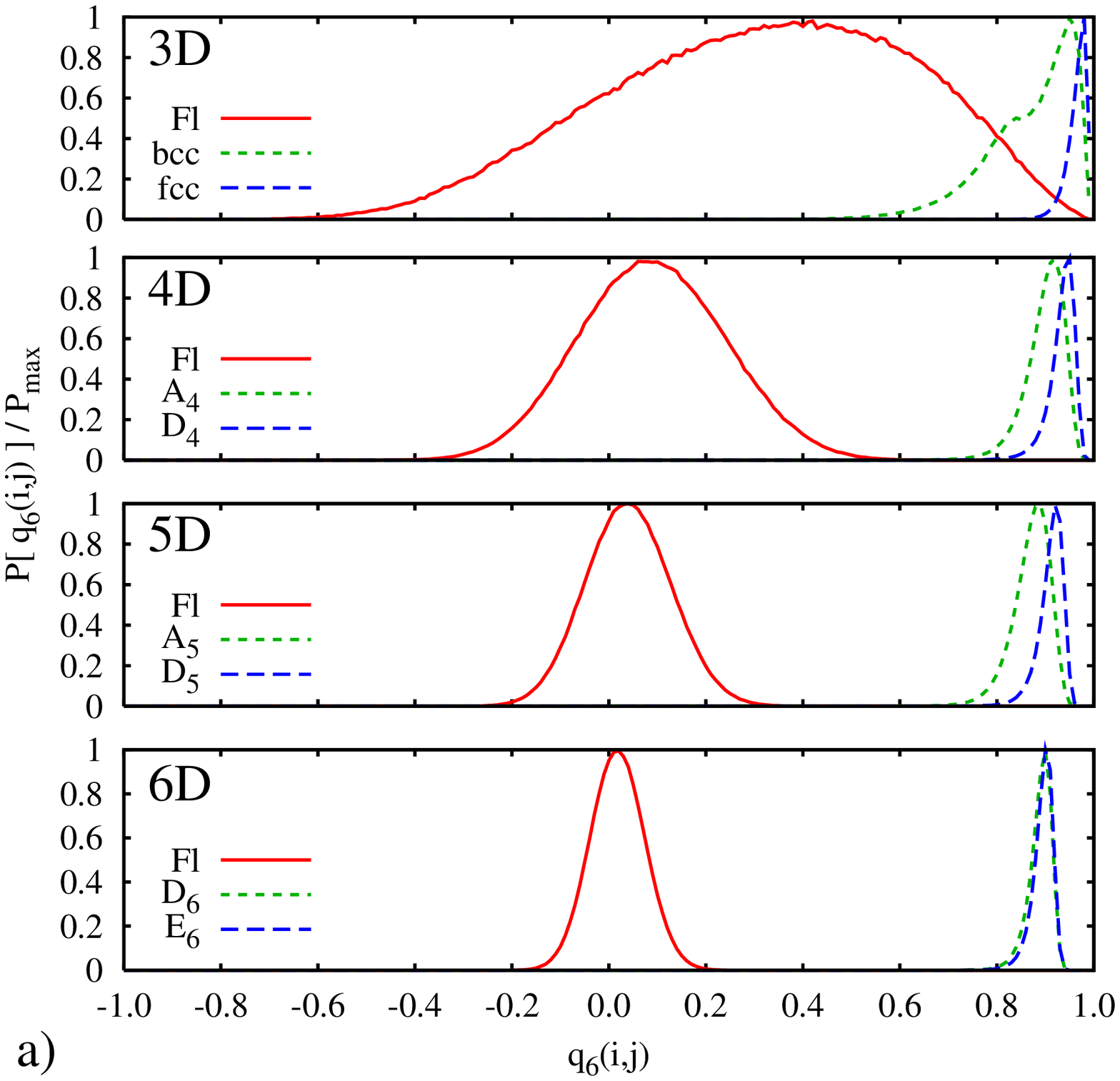}\hfill
\includegraphics[width=\columnwidth]{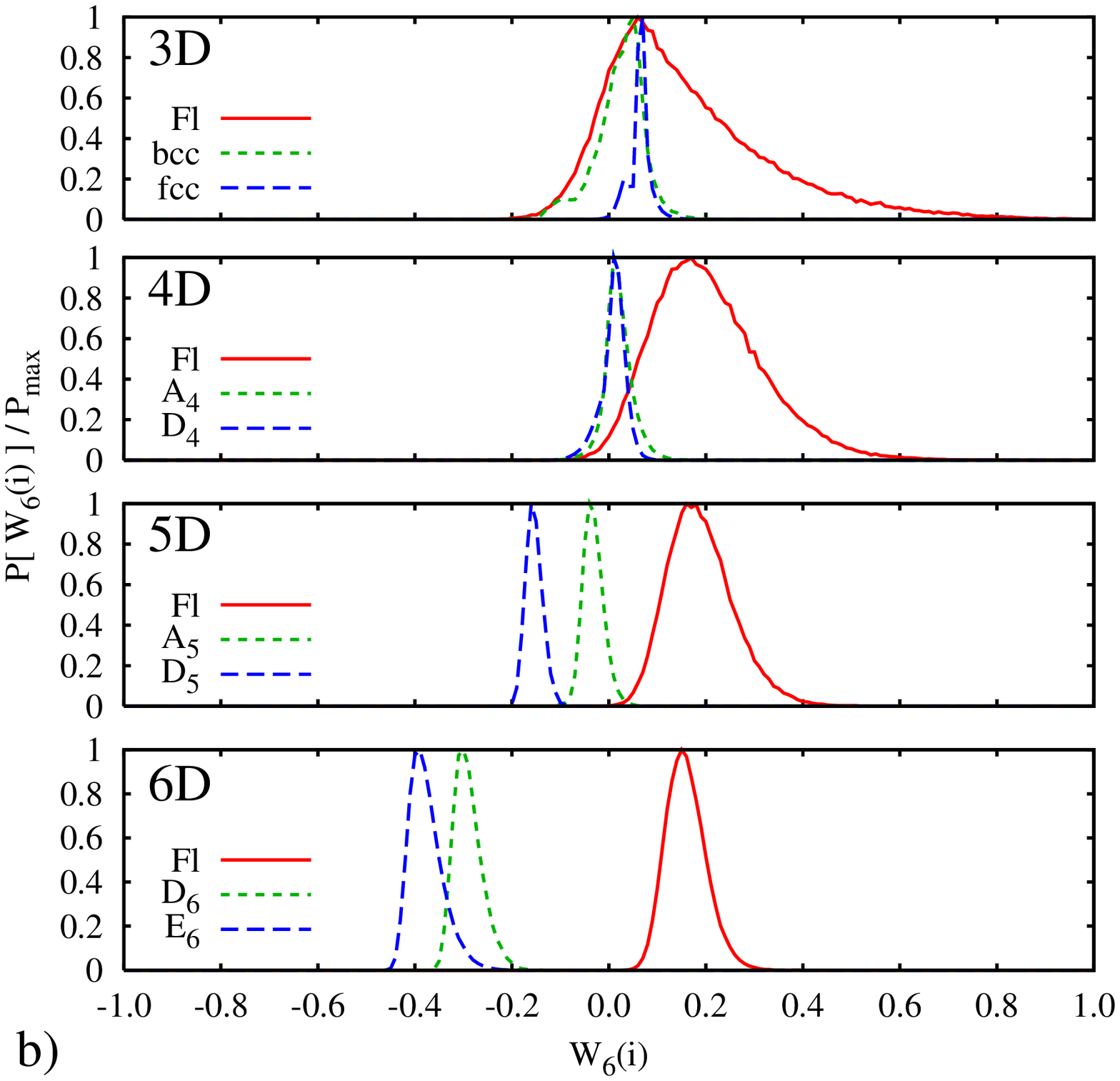}}
\caption[Order parameter]{(Color online) Distributions of second- (a) and third-order (b) $l=6$ invariants for dense crystal phases and the fluid. With increasing dimension the fluid and crystalline local bond-order parameters become increasingly distinct.}
\label{fig:bondorder}
\end{figure*}

Skoge et al., considering the radial pair distribution function $g(r)$, found that higher-order unconstrained spatial correlations vanish with increasing system dimension~\cite{skoge:2006}. In accordance to the ``decorrelation principle''~\cite{torquato:2006}, we also expect orientational correlations of order $l$, $g_l(r)$, to decay more rapidly when the dimension increases.  As seen in Figure~\ref{fig:bondcorrelator}, in 2D the hexatic signature gives rise to long-ranged orientational correlations with a length that diverges on approaching coexistence~\cite{halperin:1978,frenkel:1979}; in 3D the orientational order stretches over a couple of particle radii, but stays finite even in the supersaturated regime; and in higher dimensions the correlations keep decaying, as can be assessed from the height ratio of the second to the first peak of $g_6(r)$ in Fig.~\ref{fig:bondcorrelator}. Note that a similar behavior is observed for all $l$ considered, but $l=6$ has the advantage of capturing the crystal symmetry for all dimensions.

The authors of Ref.~\cite{skoge:2006} remarked that the number of particles counted in the first peak of $g(r)$ for supersaturated fluids matches the number of kissing neighbors in the densest known lattice phase for a given dimension. They hypothesized that disordered packings in higher dimension might thus be built of deformed crystal unit cells, in contrast to the three-dimensional case where ``icosahedral'' order was thought to dominate the packing. The distributions of local bond-order correlators, which shows how the relative crystal and fluid local orders evolve with dimensionality paint a different picture (see Fig.~\ref{fig:bondorder}). Both the second- and third-order invariants in 4D to 6D, capture no significant overlap between the liquid and the crystal local bond-order parameters, in contrast to 2D and 3D where the bond-order distributions overlap significantly~\cite{auer:2004}. Moreover, the distinction between the fluid and crystal phases increases with dimension, which suggest that the fluid and crystal local orders are just as or more distinct with increasing dimensionality, not less~\cite{footnote:2}. In 4D, where the maximally-kissing cluster 24-cell~\cite{musin:2004} is also the unit cell of the crystal, but is not a simplex-based structure, no hint of the crystal order in the fluid is captured by the bond-order distribution (Fig.~\ref{fig:bondorder})~\cite{vanmeel:2009}. For a simplex-based cluster to have as many nearest neighbors within the first peak of $g(r)$ as in the crystal, the first neighbor spheres cannot all be kissing the central sphere at the same time, but have to fluctuate in and out of the surface of the central sphere. This variety of possible configurations is what broadens the first peak of $g(r)$ and by ricochet its second peak as well~\cite{skoge:2006}. Though they are harder to illustrate geometrically, similar phenomena are expected in higher dimensions. The bond-order distribution is therefore fully consistent with a fluid structure dominated by simplex-based order, but not with the presence of deformed crystal unit cells. This clear separation in local order also suggests that for dimensions greater than three fluid configurations should be easier to distinguish from the partially crystalline or polycrystalline systems that can be observed in 3D compression studies~\cite{torquato:2000}.

\subsection{Fluid-Hard-Wall Interfacial Free Energy}
\label{sect:gammaf-w}
\begin{figure}
\center{\includegraphics[width=\columnwidth]{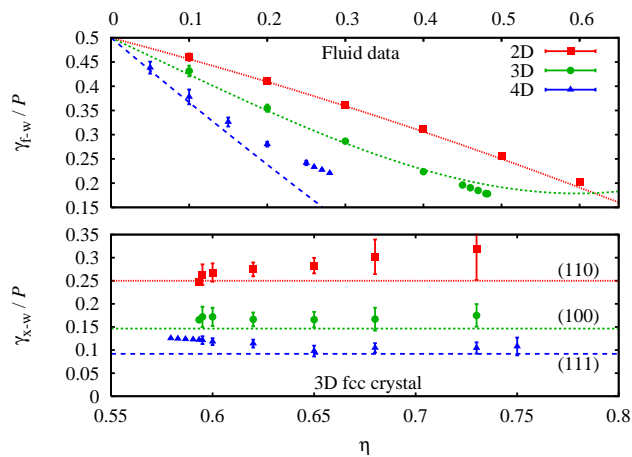}}
\caption[Fluid-hard-wall interfacial free energy]{(Color online) Top: Additional fluid volume per unit area of surface $\Delta v_A=\gamma_{\mathrm{f-w}}/P$ from Monte Carlo simulations and the Pad\'e approximant equation of state~\cite{clisby:2004,bishop:2005} (points) compared with the virial expansion of Eq.~\ref{eq:vol_virial} (lines). Bottom: Comparison of $\gamma_{\mathrm{x-w}}/P$ from published simulation data~\cite{laird:2007,heni:1999} and the Speedy equation of state~\cite{speedy:1998} (points) with the cell-like theory (see text) for different orientations of the 3D fcc crystals (lines).}
\label{fig:gammawall}
\end{figure}

The local bond-order distributions indicate that the hard-sphere fluid order resembles more the crystal order in 3D than in higher dimensions. Because of the purely entropic nature of hard-sphere systems,  microscopic geometry should have a clear thermodynamic signature on quantities such as the fluid-crystal interfacial free energy $\gamma_{\mathrm{f-x}}$. We will get back to this point in Section~\ref{sect:gammaf-x}. We first consider the behavior of the fluid-hard-wall interfacial free energy $\gamma_{\mathrm{f-w}}$, which is easier to interpret microscopically and is a limit case for the fluid-crystal geometrical frustration in all dimensions. From the fluid point of view, both the hard wall and the crystal plane exclude configurations containing spheres less than a radius away from the hard surface, but the crystal has additional free volume crevasses to explore.

The fluid-hard-wall interfacial free energy is an equilibrium quantity that, unlike the liquid-crystal interfacial free energy, is well defined at all densities where the liquid is stable.
An ideal gas has no fluid-hard-wall interfacial free energy, but the low density limit of spherical particles that exclude volume is $\gamma_{\mathrm{f-w}}=P/2+ \mathcal{O}(\rho^2)=\rho/2+\mathcal{O}(\rho^2)$, because of the $PV$ work required to exclude particles from the surface of the hard wall. At higher densities, just like a surfactant reduces the surface tension by occupying part of the interfacial area, the presence of particles at the interface reduce the entropy cost for the other particles and partly offsets the increase in opposing bulk pressure. An exact expansion gives~\cite{stillinger:1962,bellemans:1962,stecki:1978}
\begin{equation}
\gamma_{\mathrm{f-w}}=\frac{P(\rho)}{2}-B_{\Omega 2}\rho^2 - B_{\Omega 3}\rho^3-B_{\Omega 4}\rho^4-\mathcal{O}(\rho^5),
\end{equation}
where the first couple of surface virial coefficients $B_{\Omega i}$ are computed for hard spheres in all dimensions in Appendix~\ref{sect:virial}.

The virial expansion can be contrasted with the SPT and the ``mechanical'' expressions for $\gamma_{\mathrm{f-w}}$. It has already been noted that 3D SPT, captures the first term in the virial expansion exactly and is within a few percent of the next order term~\cite{stecki:1978}. This seems to be the case for all odd dimensions.  For even dimensions, though SPT captures the low-density pressure behavior correctly, it is off already for $B_{\Omega 2}$. In 2D SPT  gives $B_{\Omega 2}=\pi/8$, while the correct value is $1/3$. Even when using a precise expression for the pressure, SPT does not capture the inversion of $\gamma_{\mathrm{f-w}}^{\mathrm{coex}}$ from 3D to 4D (see Table~\ref{tab:surfacegamma}). Similarly, it has also been pointed out that the ``mechanical'' result of Kirkwood and Buff with a Fowler-type approximation for the two-point correlation~\cite{kirkwood:1949,fowler:1937} (see also Ref.~\cite[Sect. 4.4]{rowlinson:2002}) correctly captures the first surface virial coefficient in 3D~\cite{barker:1976}, and in all dimensions generally. Higher order terms contain surface effects that the simple decorrelated approximation cannot capture, which explains why it is about $60\%$ off for $B_{\Omega 3}$ in 3D, predicting  $B_{\Omega 3}=5\pi^2/96$ though the exact value is $149\pi^2/1680$. %In higher dimensions it does not even get the sign of $B_{\Omega 2}$ right.
One should therefore take quantitative $\gamma_{\mathrm{f-w}}$ predictions of those two approximations with a grain of salt.

To better understand the density dependence of $\gamma_{\mathrm{f-w}}$, we remove its trivial pressure contribution and obtain the change in fluid volume per unit of wall area
\begin{equation}
\Delta v_A(\rho)\equiv\frac{\gamma_{\mathrm{f-w}}(\rho)}{P(\rho)}.
\label{eq:gamma_vol}
\end{equation}
Using the standard pressure virial coefficients $B_i$, we expand $\Delta v_A$ to second order in density
\begin{equation}
\Delta v_A(\rho)=\frac{1}{2}-B_{\Omega 2}\rho +(B_2 B_{\Omega 2}-B_{\Omega 3})\rho^2 + \mathcal{O}(\rho^3).
\label{eq:vol_virial}
\end{equation}
For 3D hard spheres $B_{\Omega 4}$ has been evaluated numerically~\cite{stecki:1978}, so the expansion can be carried to the next order. As expected, the initial slope of $\Delta v_A(\rho)$ is always negative (see Table~\ref{tab:virial}). Also, the coefficient of $\rho^2$, which is initially negative, changes sign between 3D and 4D. Because the $\rho^2$ coefficient is small in 3D and 4D, the next order coefficient is more significant and indeed $B_{\Omega 4}$ markedly improves agreement with 3D MC results. Fig.~\ref{fig:gammawall} shows that theory and simulations results nearly coincide at low density, which validates the current simulation methodology and the latest 3D results~\cite{demiguel:2006,laird:2007}. For 3D the match with theory extends close to the coexistence limit. Only one or two additional coefficients would probably be needed to capture the full curve with simulation accuracy at coexistence.

\begin{figure}
\center{\includegraphics[width=0.6\columnwidth]{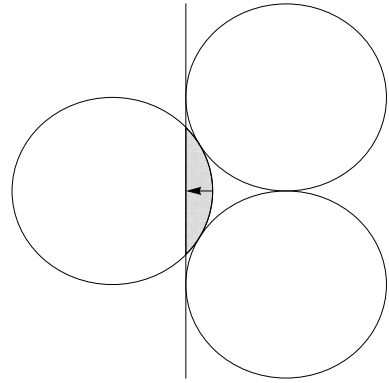}}
\caption[Fluid-hard-wall interfacial free energy]{Additional volume required for placing a wall through a simplex. The disk on the left must be pushed back sufficiently (arrow) to allow for the wall insertion (vertical line).}
\label{fig:diagram}
\end{figure}
Based on the physical interpretation of $\Delta v_A$, we expect the change in fluid volume to monotonically decay to a positive constant at high densities. By breaking the translational symmetry the introduction of a hard wall must perturb the fluid order to some extent. We also expect a plateau to develop on approaching that limit, because the high-density structure of the supersaturated fluid varies little. But it might only be possible to observe this plateau at high fluid density, in the supersaturated regime. In 2D high density reliable results are technically difficult to obtain. The possible presence of an hexatic phase at high densities commands very large system sizes, because the wall favors the local crystal order and pushes defects away from the interface. In 3D rapid wall-induced crystallization occurs before a significant degree of saturation a reached~\cite{heni:1999,fortini:2005,demiguel:2006,laird:2007,auer:2003}.
In $d>3$ however the hard interface does not accommodate a regular packing of simplexes, unlike in lower dimensions. We therefore expect heterogeneous crystallization to be sufficiently slow to study $\gamma_{\mathrm{f-w}}$ in the dense fluid regime by simulation. Though 4D $\Delta v_A$ has not yet saturated in Fig.~\ref{fig:gammawall}, the results do suggest a slowdown of the decrease. Unfortunately, the higher-dimensional system sizes required to further clarify this point are currently beyond our computational reach.

Indirect evidence nonetheless supports the saturation scenario. For crystals, saturation of $\Delta v_A$ is similar to the cell theory that was successfully applied in 3D for $\gamma_{\mathrm{f-w}}$~\cite{heni:1999}. The gap that must be opened to insert a plane along a given cut through a packed crystal is the high density limit of $2 \Delta v_A$, where the factor of two accounts for the two interfaces that are created by wall insertion. Published crystal-wall interfacial free energy $\gamma_{\mathrm{x-w}}$ simulation results for various faces of the fcc crystal~\cite{heni:1999,laird:2007}  allow to extract the corresponding $\Delta v_A(\rho)$. The lower panel of Fig.~\ref{fig:gammawall} shows that, within the error bars, the simulation results follow pretty closely this simple saturation scaling. The toy model also rationalizes the ``broken bond'' and anisotropy ratios of Ref.~\cite{heni:1999}. Because the interfacial free energy between the fcc (111) plane and a hard wall is tiled exclusively with 2D simplexes, which is similar to a 3D high density fluid-hard-wall interface, a similar $\Delta v_A$ saturation is expected. This prediction can be checked by reanalyzing the recent simulations of jamming in confinement~\cite{desmond:2009}. Desmond and Weeks considered how the close-packed density of bidisperse spheres is affected when the distance between two confining hard walls is changed. The scaling quantity $C$ they extract can be reexpressed as the additional fluid volume per wall surface area $\Delta v_A(\rho_{\mathrm{MRJ}})=C\eta_{\mathrm{MRJ}}/2.8$, where the numerical prefactor takes into account the presence of two interfaces and rescales the diameter of the larger spheres to unity. In 3D this gives $\Delta v_A(\rho_{\mathrm{MRJ}})\approx 0.054$, which is lower than the fcc (111) limit for monodisperse spheres of $0.091$. It is qualitatively expected that a bidisperse system have a reduced $\Delta v_A$, because the smaller spheres do not need to be pushed as far back from the interface as the larger spheres~\cite{amini:2008,footnote:4}.

How the high-density limit of $\Delta v_A$ changes with dimension dictates whether the free energy barrier to nucleation vanishes or not in high dimensions. From the geometrical frustration analysis, we have identified the simplex as the dominant geometrical structure in high density fluids. It is therefore reasonable to expect that the dominant structure near a hard interface be a truncated simplex. Obviously the fluid interface does not exclusively contain truncated simplexes, so we further assume that simplexes are representative of the interface, and thus dominate the volume loss. The high density limit of $\Delta v_A$ is then the distance that a vertex sphere needs to be pushed out from the other spheres in the simplex for a tangent plane to be inserted. For 2D, this scenario is depicted in Fig.~\ref{fig:diagram} and the details of the general calculation are given in Appendix~\ref{sect:wallvolume}. Within this approximation $\Delta v_A$ monotonically increases with dimension, but is bounded by $1/2-\sqrt{2}/4\approx 0.146$. In high dimensions, $\gamma_{\mathrm{f-w}}^{\mathrm{coex}}$ does not therefore asymptotically vanish, but increases, due to the growing $P^{\mathrm{coex}}$.

\subsection{Fluid-Crystal Interfacial Free Energy}
\label{sect:gammaf-x}
\begin{table}
\begin{ruledtabular}
\begin{tabular}{|c|c|c|c|c|c|}
  d & $\gamma_{\mathrm{f-x}}^{\mathrm{coex}}$ & $ \gamma_{\mathrm{f-w}}^{\mathrm{coex}}$ & $ {\rm SPT}_{1}  \gamma_{\mathrm{f-w}}^{\mathrm{coex}}$ & $ {\rm SPT}_{2} \gamma_{\mathrm{f-w}}^{\mathrm{coex}}$&$ \gamma_{\mathrm{f-x}}^{\mathrm{coex}}/ \gamma_{\mathrm{f-w}}^{\mathrm{coex}}$\\
    \hline
  3 & 0.557~\cite{laird:2007} & 1.98(1) & 2.30~\cite{heni:1999} & 1.75 &0.28  \\
  4 & 1.0 & 1.96(2) & 2.53~\cite{finken:2001} & 1.89 &0.53   \\
  5 &  - & - & 2.93 & 2.07 &  -\\
  6 &  - & - & 4.05 & 3.24 & - \\
\end{tabular}
\end{ruledtabular}
\caption[Coexistence parameters]{Fluid-crystal and fluid-hard-wall interfacial free energies at coexistence from simulations and SPT~\cite{heni:1999,finken:2001}. The column SPT$_{1}$  is the SPT interfacial tension with SPT pressure, and the column SPT$_{2}$ is the SPT interfacial tension calculated using the MC pressure.} \label{tab:surfacegamma}
\end{table}
Our previous crystallization study of 4D hard spheres gives values for $\gamma_{\mathrm{f-x}}$ that increase with fluid supersaturation (or, equivalently, pressure)~\cite{vanmeel:2009}. These interfacial free energies are more than twice as large as for 3D hard spheres at comparable supersaturation~\cite{auer:2001}. Even taking into account the slightly different interfacial densities, the gap is large, justifying our earlier claim of increased geometrical frustration in 4D~\cite{vanmeel:2009}. To allow for a more quantitative comparison, we consider here the system at coexistence, where the equilibrium fluid-crystal interfacial free energy $\gamma_{\mathrm{f-x}}^{\mathrm{coex}}$ is unambiguously defined.

Crystallization-derived $\gamma_{\mathrm{f-x}}$ for 3D hard spheres have been corrected for finite-size effects to obtain $\gamma_{\mathrm{f-x}}^{\mathrm{coex}}$~\cite{cacciuto:2003}. The effective critical cluster sizes contain at most a few spherical shells and the resulting strongly curved crystallite leads to a large internal Laplace pressure. The resulting loss of free volume per particle is compounded by the relatively large compressibility of higher-dimensional crystals near coexistence, increasing the free-energy cost of forming the interface. Just like $\gamma_{\mathrm{f-w}}$ increases with  density, however, the non-equilibrium fluid-crystal interfacial free energy for supersaturated fluids should be larger than $\gamma_{\mathrm{f-x}}^{\mathrm{coex}}$, due to the overall increase in bulk pressure. Because $\Delta v_A$ has probably not yet saturated, it is difficult to get a precise estimate of this effect, but it is of the right magnitude to explain the change of $\gamma_{\mathrm{f-x}}$ with pressure in both 3D and 4D. Which of the crystallite finite-size or the increase in fluid pressure dominates the non-equilibrium $\gamma_{\mathrm{f-x}}$ behavior cannot be resolved here. But to first order, through the use of the Tolman ``ansatz'', they both suggest that the interfacial free energy depends linearly on supersaturation $\Delta\mu$~\cite{tolman:1949}. This scaling, though microscopically inaccurate and therefore only used on an \emph{ad hoc} basis~\cite{talanquer:1995}, showed a relative success in 3D~\cite{cacciuto:2003,zykova-timan:2008}, and gives $\gamma^{\mathrm{coex}}_{\mathrm{f-x}}\approx 1.0$ for 4D hard spheres. In spite of its crudeness, the result is sufficiently precise to interpret the thermodynamic consequences of geometrical frustration, because the equivalent quantities in 3D are known with high  accuracy~\cite{davidchack:2006,laird:2007}.

To a first approximation one expects the fluid-crystal interfacial free energy to scale linearly with the fluid-hard-wall quantity if geometrical frustration were constant in all dimensions, because the depth of the interface remains of the order of the particle dimension. Yet comparing the ratio of fluid-crystal to fluid-hard-wall interfacial free energies in 3D and 4D in Table~\ref{tab:surfacegamma} shows this not to be universally the case. To understand the origin of the difference we consider how the interfacial free energy decreases upon changing the hard wall to a crystal interface. The core of the reduction comes from two sources: the crystal planes near the fluid interface have more free volume than those in the bulk and the interfacing fluid requires less $\Delta v_A$ than next to hard wall. The interfacial picture suggested by DFT, where in 3D the bulk of the interfacial free energy comes from the set of three or four layers that form the interface, is consistent with both scenarios~\cite{curtin:1989}.
In 3D, the crystal relaxation due to the additional free volume for interfacial particles has recently been considered the dominant effect~\cite{laird:2001}. But an older toy model by Spaepen ascribes a significant contribution to the fluid~\cite{spaepen:1975,spaepen:1976}.Though our results cannot resolve quantitatively the relative contribution of the two scenarios, they at least indicate that both are of comparable magnitude~\cite{footnote:1}. If the crystal modulation alone explained the reduction of the interfacial free energy from the hard wall limit, it should be similar if not greater in 4D than in 3D, because the reduced $P^{\mathrm{coex}}$ in 4D allows for more wandering of the interfacial crystalline particles. Yet the fluid-crystal interfacial free energy is both proportionally and absolutely larger in 4D than in 3D. The further reduction of $\gamma_{\mathrm{f-x}}^{\mathrm{coex}}$ in 3D must therefore arise from the relatively good geometrical match between the fluid and the crystal order at the interface. The crystal surface allows for more additional microstates in the spacing between the interfacial crystal spheres in the 3D fcc lattice than in the 4D $D_4$ lattice. Based on this observation and the almost complete bond-order parameter mismatch between the 4D crystal phases and the fluid it is reasonable to describe 4D as completely geometrically frustrated and 3D as only partially so.

What about higher dimensions? Because the local bond-order mismatch is already very pronounced in 4D (see Sect.~\ref{sect:bond-orderdist}), it is hard to imagine that higher dimensions could show any more geometrical frustration than 4D does. Once the fluid and crystal order distributions stop overlapping significantly, the fluid simply does not spontaneously form structures that easily anchor to the crystal interface. We therefore expect the coexistence fluid-crystal interfacial free energy to scale linearly  with $\gamma_{\mathrm{f-w}}^{\mathrm{coex}}$, but only for $d\geq4$.

Let us consider for a moment the impact of these observations on 3D polydisperse systems. Increasing polydispersity normally decreases $\Delta v_A(\rho)$, as discussed above. The increase of $\gamma_{\mathrm{f-x}}^{\mathrm{coex}}$ with polydispersity thus occurs because $P^{\mathrm{coex}}$ increases faster~\cite{bolhuis:1996} than $\Delta v_A(\rho^{\mathrm{coex}})$ decreases. The higher $P^{\mathrm{coex}}$ and $\gamma_{\mathrm{f-x}}^{\mathrm{coex}}$ then lead to a higher free energy barrier and to a more rapid increase in non-equilibrium $\gamma_{\mathrm{f-x}}$ with supersaturation~\cite{auer:2001b}. The curvature of the crystal nucleus appears to be a marginal contribution~\cite{cacciuto:2004}.

\subsection{Classical Nucleation Theory}
\label{sect:CNT_discuss}
In the end, what do these results imply for the crystallization barrier? The geometrical interpretation for $\gamma_{\mathrm{f-w}}$ above as well as the connection between $\gamma_{\mathrm{f-w}}$ and $\gamma_{\mathrm{f-x}}$ suggested by geometrical frustration permit certain predictions. Because $\Delta v_A(\rho)$ does not vanish in high dimensions and because $P^{\mathrm{coex}}$ increases with dimensionality, due to the increasing inefficiency of lattice packings to fill space compared to simplex-based fluids, we expect $\gamma_{\mathrm{f-x}}^{\mathrm{coex}}$ to grow with dimensionality. In the denominator of Eq.~\ref{eq:barrtrend} the crystal density increases with dimension, but the cell model provides a lower bound for the pressure contribution to $\gamma_{\mathrm{f-x}}$ that also scales linearly with density and with a larger prefactor. Because the geometrical prefactor scales like $d^{d/2}$, the nucleation barrier of monodisperse hard spheres thus increases with dimension. Higher-dimensional crystallization becomes ever rarer, which explains the surprising stability of supersaturated monodisperse hard sphere fluids observed in simulation~\cite{skoge:2006,vanmeel:2009}.

\section{Conclusion}
The modern understanding of geometrical frustration in hard sphere fluids considers how much space would need to be curved to allow for maximally dense simplex-based structures to form lattices~\cite{sadoc:1999,nelson:2002,tarjus:2005}. Though ultimately this lack of curvature is the reason why lattices in Euclidean space cannot be simple assemblies of simplexes, it lacks a direct dynamical mechanism to prevent crystallization. This study has made more precise the role of geometrical frustration in increasing the interfacial free energy between the fluid and the crystal in monodisperse and polydisperse systems.
We have argued that fluid-hard-wall interfacial free energy is an upper bound for the fluid-crystal interfacial equivalent, whose scaling behavior is easier to model. We have also argued that based on the poor overlap between the local fluid and crystal orders $\gamma_{\mathrm{f-x}}$ and $\gamma_{\mathrm{f-w}}$ should remain of the same order of magnitude in high dimension.
Put together, these elements allow us to predict that in high dimensions the free energy barrier to crystallization is much larger than in 3D, and therefore crystallization is much rarer than in 3D. The crystal thus only marginally impacts a supersaturated fluid dynamics in high dimensions, unless a different type of phase transition arises~\cite{frisch:1999}.
For the regime where that is not the case, higher dimensional spheres are an interesting model in which to study phenomena that are  ambiguous in 3D, such as jamming, as we saw above, and glass formation~\cite{charbonneau:2009}.

Why is crystallization then so common in 3D hard spheres? First, the crystal is sufficiently efficient at filling space for the coexistence pressure to be relatively low. Second, the overlap of the fluid and crystal order-parameter distributions is significant in 3D, which results in only a moderate geometrical frustration. Three-dimensional space is so small that all possible cluster organizations, including the cubic crystal unit cells, are frequently observed in the fluid, which limits the extent of geometrical frustration. Bernal and many after him have indeed observed just about any small polyhedron in hard sphere fluids~\cite{bernal:1959}. Though the simplex-based fluid order is preferred, other ordering types are not far off and can easily be accommodated. Geometrical frustration in 3D monodisperse hard spheres is thus only partial, because even the crystal unit cell is sufficiently ``liquid-like''.

Finally, our study has shown that many microscopic details of the interfacial free energy of even the simplest of systems are mis- or incompletely understood. Any hope to get a grasp of and control homogeneous and heterogeneous crystallization in more complex systems is contingent upon having a better understanding of these fundamental issues.

\begin{acknowledgments}
We thank S.~Abeln, K.~Daniels, D.~Frenkel, B.~Laird, M.~Maggioni, H. Makse, R.~Mossery, B.~Mulder, M.~Stern, G.~Tarjus, P. Wolynes, and F. Zamponi for discussions and suggestions at various stages of this project. The work of the FOM Institute is part of the research program of FOM and is made possible by financial support from the Netherlands Organization for Scientific Research (NWO).
\end{acknowledgments}

\appendix

\section{Lattice Generating Matrices}
\label{sect:genmatrix}
The canonical lattice representations are often given in a form that simplifies the notation~\cite[Chap. 4]{conway:1988}, but does not necessarily allow for a convenient physical construction for simulation purposes. Moreover, the choice of generating lattice vectors should keep the size of the simulation box commensurate with the crystal unit cell to a minimum. $D_d$ packings are checkerboard lattices, which are algorithmically simple to generate. $A_4$ and $A_5$ are dense packings with generating matrices
\begin{equation}
\left(
     \begin{array}{cccc}
       0                & 1     & 0             & 0 \\
       0                & 1/2   & \sqrt{2}/2    & 1/2 \\
       0                & 0     & 0             & 1 \\
       \sqrt{10}/4      & 0     & 1/4           & 1/2
     \end{array}
   \right)
\end{equation}
and
\begin{equation}
\left(
     \begin{array}{ccccc}
       0           & 1    & 0           & 0             & 0 \\
       0           & -1/2 & \sqrt{3}/2  & 0             & 0 \\
       0           & 0    & -\sqrt{3}/3 & \sqrt{6}/3    & 0 \\
       \sqrt{10}/4 & 0    & 0           & \sqrt{6}/4    & 0 \\
       \sqrt{10}/5 & 0    & 0           & 0             & \sqrt{15}/5
     \end{array}
   \right)
\end{equation}
respectively. $E_6$ is a cut through the $E_8$ generalization of the diamond lattice, for which we use the generating matrix
\begin{equation}
\left(
  \begin{array}{cccccc}
   0		&  0		&  0		&  \sqrt{3}   & 0           & 0 \\
   0		&  0	  &  0		&  0          &  \sqrt{3}   & 0 \\
   1		&  1		&  1		&  0		      & 0			      & 0 \\
   1		& -1/2  & -1/2  &  0		      & -\sqrt{3}/2	& -\sqrt{3}/2 \\
  -1/2	&  1		& -1/2  & -\sqrt{3}/2	& 0			      & -\sqrt{3}/2 \\
   1/2	&  1/2  &  1/2  &  \sqrt{3}/2	&  \sqrt{3}/2	&  \sqrt{3}/2
  \end{array}
\right).
\end{equation}

From these generating matrices we can construct a unit cell commensurable with our hyper-rectangular simulation box. The sides of the unit cell $u_i$ are obtained by linear combinations of the matrix' row vectors such that only one non-zero element remains. The lattice sites located within the unit cell borders then correspond to the particle positions within the unit cell. Following this recipe, our $A_4$ unit cell has relative side dimensions $l = ( \sqrt{10}, 1, \sqrt{2}, 1 )$ with $n_u = 8$ particles in the unit cell, $A_5$ yields $l = ( \sqrt{10}, 1, \sqrt{3}, \sqrt{6}, \sqrt{15} )$ with $n_u = 120$, and $E_6$ yields $l = ( 3, 3, 3, \sqrt{3}, \sqrt{3}, \sqrt{3} )$ with $n_u = 24$.

\section{Third-Order Invariant Polynomials}
\label{sect:thirdorder}
For reference, we provide two common 3D third-order rotationally-invariant polynomials $\tilde{w}^d_m(a,b,c)$
for l=4 and l=6, expressed in inner products as outlined in Sec.~\ref{sec:metho-order}. The simple 4D $l=4$
polynomial is also given.
\begin{align}
&\begin{aligned}
\tilde{w}^3_4=&
1+{\frac {3700}{109}}abc+\frac{8575}{109}a^2b^2c^2
-5(a^2+b^2+c^2)\\ &+\frac{490}{109}(a^4+b^4+c^4)-\frac{4900}{109}(abc^3+a{b}^{3}c+a^3bc)\\
&+\frac{875}{109}(a^2b^2+b^2c^2+a^2c^2),
\end{aligned}
\end{align}

\begin{align*}
\tilde{w}^3_6=&
1+\frac{47187}{262}abc+{\frac {369117}{262}}a^2b^2c^2
+{\frac {586971}{262}}a^3b^3c^3\\
&-\frac{21}2(a^2+b^2+c^2)+\frac {7371}{262}(a^4+b^4+c^4)\\
&-\frac{2541}{131}(a^6+b^6+c^6)
+{\frac {2646}{131}}(a^2b^2+b^2c^2+a^2c^2)\\
&-{\frac {145089}{262}}(a^3bc+ab^3c+abc^3)\\
&+{\frac{53361}{131}}(a^5bc+ab^5c+abc^5)\\
&-{\frac {480249}{262}}(a^4b^2c^2+a^2b^4c^2+a^2b^2c^4)\\
&+{\frac {189189}{262}}(ab^3c^3+a^3bc^3+a^3b^3c)\\
&-\frac{4851}{262}(a^2b^4+a^2c^4+b^2c^4+b^4c^2+a^4c^2+a^4b^2),
\end{align*}
and
\begin{align}
&\begin{aligned}\tilde w^4_4=&1+\frac{264}5abc+\frac{768}5{a}^{2}{b}^{2}{c}^{2}
-6(a^2+b^2+c^2)
\\&+\frac{32}5(a^4+b^4+c^{4})+\frac{48}5({b}^{2}{c}^{2}+{a}^{2}{c}^{2}+{a}
^ { 2 } { b } ^ { 2 } )
\\&
-\frac{384}5(a^3bc+a{b}^{3}c+ab{c}^{3}).
\end{aligned}
\end{align}

\section{Statistical Mean-Field Theory of Jamming}
\label{sect:statjam}
\begin{table*}
\begin{ruledtabular}
\begin{tabular}{|c|c|c|c|}

  % after \\: \hline or \cline{col1-col2} \cline{col3-col4} ...
  $d$ & $2^dV^*(c)/V_d$ & $S^*(c)$ & $S_{\mathrm{occ}}$ \\
  \hline
  4 & $\frac{2}{3 c^2 \pi } \left[\sqrt{-1+c^2} \left(16+2 c^2+3 c^4\right)+\right.$ & $2 \pi  \left[\arccos{(1/c)}- \sqrt{1/c^2-\frac{1}{c^4}}\right]$ & $\frac{\pi}{6}\left(2\pi-3\sqrt{3}\right)$ \\
    & $\left.3 c^2 \left(-8+c^4\right) \arccos(1/c)\right]$ & &\\
  5 & $-16 - 5/c^3 + 20/c + c^5$ & $2 \pi^2\frac{(-3 + 1/c^2 + 2 c)}{3 c}$ & $2\pi^2 \left(\frac{2}{3} - \frac{3 \sqrt{3}}{8}\right)$ \\
  6 & $\frac{2}{15 c^4 \pi} \left[\sqrt{-1 + c^2} \left(-192 + 624 c^2 + 8 c^4 + 10 c^6 + 15 c^8\right) + \right.$ & $\frac{\pi^2}{12}\left[12 \arccos{(1/c)} - \right.$&  $\frac{\pi^2}{24}\left(-7 \sqrt{3} + 4 \pi\right)$\\
    & $\left.
   15 c^4 \left(-32 + c^6\right) \arccos{(1/c)}\right]$ & $\left.8 \sin\left(2 \arccos{(1/c)}\right) + \sin\left(4 \arccos{(1/c)}\right)\right]$ &\\
\end{tabular}
\end{ruledtabular}
\caption{Volume functions for the statistical mean-field jamming analysis of Ref.~\cite{song:2008} in higher dimensions.}
\label{table:sstar}
\end{table*}
Using the notation, approach, and approximations of Ref.~\cite{song:2008}, we obtain the theoretical predictions of $\eta_{\mathrm{MRJ}}^{\mathrm{stat}}$ for higher dimensions. For mechanically stable configurations, the bulk and contact contributions to the cumulative probability that the coordinates of all spheres $j$ obey $r_j/\cos{\theta_j}>c$ is
\begin{equation*}
P_>(c)\approx \exp{\left(\frac{-2^d V^*(c)}{V_d (1/\eta_{\mathrm{MRJ}}^{\mathrm{stat}} -1)} +\frac{-2d S^*(c)}{S_{d-1}/2 + S_{d-2}S_{\mathrm{occ}}}\right)}
\end{equation*}
with (see Table~\ref{table:sstar})
\begin{align}
V^*(c)&=\int\Theta(c-r/\cos\theta)d\mathbf{r}\\
S^*(c)&=\int\delta(r-1)\Theta(c-r/\cos\theta)d\mathbf{r}\\
S_{\mathrm{occ}}&=\int_0^{\pi/6}\sin^{d-2}\theta d\theta.
\end{align}
A self-consistent solution to 
\begin{equation}
1/\eta_{\mathrm{MRJ}}^{\mathrm{stat}}-1=-\int_1^{\infty}(c^d-1)\frac{dP_>(c)}{dc}dc.
\end{equation}
obtained numerically gives $\eta_{\mathrm{MRJ}}^{\mathrm{stat}}$.

\section{Hard-Sphere Surface Virial Coefficients}
\label{sect:virial}
Using the notation of Ref.~\cite{stecki:1978}, we calculate the virial
corrections to the interfacial free energy between a hard sphere fluid and a
hard wall. The first two coefficients of the expansion
\begin{align}
B_{\Omega 2}&=I^d_{21}/2\label{BOd}\\
B_{\Omega 3}&=I^d_{31}/2-(2I^d_{32}+I^d_{33})/6\label{BOt}
\end{align}
are given in Table~\ref{tab:virial}
for $d\leq 10$.\begin{table}
\begin{ruledtabular}
\begin{tabular}{|c|c|c|}
  $d$ & $B_{\Omega2}$ & $ B_{\Omega 3}$\\
    \hline
  2& $1/3$ & $\frac{4}{9}\pi-\frac{9\sqrt{3}}{20}$\\
  3 & $\pi/8$ & $\frac{149}{1680}\pi^2$\\
  4 &  $\frac{2}{15}\pi$ & $\frac{4}{45}\pi^3-\frac{143\sqrt{3}}{1400}\pi^2$
\\
  5 &  $\pi^2/24$ & $\frac{5375}{532224}\pi^4$  \\
  6 &  $\frac{4}{105}\pi^2$ &
$\frac{8}{945}\pi^5-\frac{1828\sqrt{3}}{175,175}\pi^4 $  \\
  7 &  $\pi^3/96$ & $\frac{266,977}{415,135,720}\pi^6$  \\
  8 &  $\frac{8}{945}\pi^3$ & $\frac{4}{8505}\pi^7 -\frac{144,213
\sqrt{3}}{238,238,000}\pi^6 $  \\
  9 &  $\pi^4/480$ & $\frac{1,127,359,391}{42,908,324,659,200}\pi^8 $\\
  10 &  $\frac{16}{10395}\pi^4$ & $\frac{8}{467,775}\pi^9-\frac{3,945,351
\sqrt{3}}{174,271,097,000}\pi^8 $  \\
\end{tabular}
\end{ruledtabular}
\caption[Surface virial coefficients]{Surface virial coefficients for hard
spheres in low dimensions.} \label{tab:virial}
\end{table}

A convenient way to compute
these coefficients uses the volume $V^{\mathrm{cap}}_d(h)$ of a spherical cap of
height $h$ on the unit sphere $x_1^2+\cdots+x_d^2\leq 1$.  This cap is obtained
as the portion lying above the plane $x_d=1-h$.

We can then compute the quantities of  Equations (\ref{BOd}) and (\ref{BOt}) by
use of the integrals
% \begin{widetext}
\begin{align}
I^d_{21}=&\int_0^1 V_d^{\mathrm{cap}}(z_1)dz_1,\\
I^d_{31}=&\int_0^1 [V_d^{\mathrm{cap}}(z_1)]^2 dz_1,\\
I^d_{32}=&2 S_{d-2}\int_0^1\!\!
\int_0^{r_2}\!\!\!\int_0^{\arccos\frac{z_1}{r_2}}
\!\! V_d^{\mathrm{cap}}\left(1-\frac{r_2}{2}\right)\times\nonumber\\ & r_2^{d-1}\sin^{d-2}{\theta}d\theta
dz_1dr_2,\\
I^d_{33}=&-I^d_{32}/2.
\end{align}
% \end{widetext}
The last result is obtained by noticing that the mirror image across the
$z=0$ plane of a configuration is also a valid configuration and completes the
lens formed by the addition of two spherical caps. The prefactor accounts for the double
counting.

To compute $V^{\mathrm{cap}}_d(h)$, we use the spherical coordinates system $x_d=r\cos\theta$ and
$\sqrt{x_1^2+\cdots+x_{d-1}^2}=r\sin\theta$. Then the cutting plane $x_d=1-h$
intersects the sphere of radius $1$ at angle
\[\theta_h\equiv\arccos(1-h).\] Using the notation $S_{d-2}$
introduced in Section~\ref{sec:generalized_cnt} for
the surface of a $d-2$-dimensional sphere in $d-1$-dimensional space, we compute
\begin{align*}
V^{\mathrm{cap}}_d(h)&=S_{d-2}\int_0^{\theta_h}
\int_{\frac{1-h}{\cos(\theta)}}^{1}r^{d-1}\sin^{d-2}\theta dr d\theta \\
&=\frac{S_{d-2}}{d}\int_0^{\theta_h}\!\!\sin^{d-2}(\theta)-\frac{
(1-h)^d\sin^{d-2}(\theta)}{\cos^d(\theta)}d\theta
\end{align*}
Recall that for $J_d(x)\equiv \int \sin^{d}(x) dx$, we have
\[J_{2m}(x)=\frac{(2m)!\, x}{2^{2m}(m!)^2}
-\cos(x)\sum_{i=0}^{m-1}\frac{(2m)!(i!)^2\sin^{2i+1}
(x)}{2^{2m-2i}(2i+1)!(m!)^2}\]
while
\[J_{2m+1}(x)=-\cos(x)\sum_{i=0}^m\frac{2^{2m-2i}(m!)^2(2i)!\sin^{2i}(x)}{
(2m+1)!(i!)^2}.\]
Using this information, we find
\[V_d^{\mathrm{cap}}(h)=\frac{S_{d-2}}{d}\Bigl(
J_{d-2}(\theta_h)-J_{d-2}(0)
-\frac{(1-h)^d\tan^{d-1}(\theta_h)}{d-1}
\Bigr).\]
Note that $\sin(\theta_h)=\sqrt{h(2-h)}$ and therefore
\begin{widetext}
\[J^{h}_{d-2}\equiv J_{d-2}(\theta_h)-J_{d-2}(0)=\begin{cases}
\frac{2^{d-3}(\frac{d-3}{2}!)^2}{(d-2)!}\left(1-(1-h)
\sum_{i=0}^{(d-3)/2}\frac{(2i)!(h(2-h))^{i}}{4^{i}(i!)^2}\right)
          ,& \text{ if }d\text{
odd};\\	
\frac{(d-2)!}{2^{d-2}(\frac{d-2}{2}!)^2}\left(\arccos(1-h)-(1-h)\sum_{i=0}^{
(d-4)/2 } \frac{4^{i}(i!)^2(h(2-h))^{i+1/2}}{ (2i+1)!}\right),& \text{
if }d\text{ even}.
         \end{cases}\]
\end{widetext}

\section{Height Of The Cap Created By The Insertion Of A Wall}
\label{sect:wallvolume} In this appendix, we compute the height $h$ of the
spherical cap created by inserting a wall in a densely packed
simplex, which we use to approximate $\Delta v_A$. To simplify the computation, it is convenient to think of $\R^d$ as sitting in
$\R^{d+1}$ as the hyperplane  $x_1+\cdots+x_{d+1}=1/\sqrt{2}$.
Let $e_1,\ldots,e_{d+1}$ stand for the usual coordinate
basis vector. We put hard spheres centered at $e_i/\sqrt{2}$. These spheres are all at a distance
$1$ from each other and form a $d$-dimensional simplex.
When inserting a wall tangent to the spheres centered at
$e_1/\sqrt{2},\ldots,e_d/\sqrt{2}$, we cut a spherical cap from the sphere
centered at $e_{d+1}/\sqrt{2}$.  We wish to compute the height of this cap.

Let \[P_d\equiv\frac{e_1+\cdots+e_d}{d\sqrt{2}}\] be the center of mass of the
first $d$ balls.
Since $P_d$ is at a distance  $\sqrt{\frac{d+1}{2d}}$ from $e_{d+1}/\sqrt{2}$,
the wall being inserted is at a distance $\sqrt{\frac{d+1}{2d}}-\frac12$
from $e_{d+1}/\sqrt{2}$.  So the cap has height
$\frac12-\bigl(\sqrt{\frac{d+1}{2d}}-\frac12\bigr)=1-\frac{1}{\sqrt{2}}\sqrt{1+
\frac1d}$.  Dividing this result by two for the two interfaces that are created by inserting a wall and
taking the high-dimensional limit, we obtain  \[\Delta v_A\approx\frac12(2-\sqrt2)\simeq 0.146447.\]

\bibliography{text}

\begin{thebibliography}{88}
\bibitem[{foo({\natexlab{b}})}]{footnote:2}
\bibinfo{note}{In the trivial cases where the $l$-order invariants are
  incompatible with the crystal symmetry, the bond-order distribution is
  centered around zero and overlaps with the fluid
  distribution~\cite{tenwolde:1996}. The difference between the fluid and
  crystal local orders should then be assessed by other $l$-order invariants .}
\bibitem[{foo({\natexlab{d}})}]{footnote:1}
\bibinfo{note}{A more definitive separation of various contributions to
  $\gamma_{\mathrm{f-w}}$ could be achieved by considering the interfacial free
  energy of hard but patterned interfaces on approaching the coexistence
  regime. We leave this question to a future enquiry.}
\end{thebibliography}
\end{document}